\documentclass[aps,pra,twocolumn,superscriptaddress]{revtex4-1}

\usepackage{amsmath,bm,mathtools}
\usepackage{amstext}
\usepackage{epsfig}
\usepackage{xcolor}
\usepackage{subfig}
\usepackage{graphicx}
\usepackage{multirow}
\usepackage{array}
\usepackage{tikz,pgfplots}
\usepackage{tabularx, booktabs}
\usepackage{amssymb}

\definecolor{dgreen}{rgb}{0,.5,0}
\definecolor{dred}{rgb}{.7,.0,.0}

\newcommand{\etal}{{\it et al.}}

\newcommand{\dU}[1]{\ensuremath{\dfrac{\partial #1}{\partial U}}}
\newcommand{\du}[1]{\ensuremath{\dfrac{\partial #1}{\partial u}}}

\newcolumntype{Y}{>{\centering\arraybackslash}X}

\newcommand{\be}{\begin{eqnarray}}
\newcommand{\ee}{\end{eqnarray}}

\begin{document}

\title{
Multiple impurities and combined local density approximations in Site-Occupation Embedding Theory
}

\author{Bruno Senjean}
\thanks{Corresponding author}
\email{senjean@unistra.fr}
\affiliation{Laboratoire de Chimie Quantique,
Institut de Chimie, CNRS/Universit\'{e} de Strasbourg,
4 rue Blaise Pascal, 67000 Strasbourg, France}

\author{Naoki Nakatani}
\affiliation{Department of Chemistry,
Graduate School of Science and Engineering,
Tokyo Metropolitan University,
1-1 Minami-Osawa, Hachioji,
Tokyo 192-0397, Japan}
\affiliation{Institute for Catalysis,
Hokkaido University,
N21W10 Kita-ku, Sapporo,
Hokkaido 001-0021, Japan}

\author{Masahisa Tsuchiizu}
\affiliation{Department of Physics,
Nara Women's University,
Nara 630-8506, Japan}

\author{Emmanuel Fromager}
\affiliation{Laboratoire de Chimie Quantique,
Institut de Chimie, CNRS/Universit\'{e} de Strasbourg,
4 rue Blaise Pascal, 67000 Strasbourg, France}


\begin{abstract}

Site-occupation embedding theory
(SOET) is an in-principle-exact multi-determinantal extension of
density-functional theory for model Hamiltonians.
Various extensions of recent developments in SOET [Senjean {\etal},
Phys. Rev. B 97, 235105 (2018)] are explored in this work. An important
step forward is the generalization of the theory to multiple impurity
sites. We also propose a new single-impurity density-functional
approximation (DFA) where the
density-functional impurity correlation energy of the two-level (2L)
Hubbard system is
combined with the Bethe ansatz local density approximation (BALDA) to
the full correlation energy of the (infinite) Hubbard model. 
In order to test the new DFAs, the impurity-interacting wavefunction has
been computed
self-consistently with the
density matrix renormalization group method (DMRG).
Double
occupation and per-site energy expressions have been derived and
implemented in the one-dimensional case. A detailed analysis of the results is presented, with a
particular focus on the errors induced either by the energy functionals
solely or
by the self-consistently converged densities. Among all the DFAs
(including those previously proposed), the combined
2L-BALDA is the one that performs the best in all correlation and
density regimes. Finally, extensions in new directions, like a partition-DFT-type
reformulation of SOET, a projection-based SOET approach, or the combination of SOET with Green functions,
are briefly discussed as a perspective.  

\end{abstract}

\maketitle

\section{Introduction}\label{sec:intro}

The accurate and low-cost description of strongly correlated materials 
remains one of the most challenging task in electronic structure theory
As highly accurate wavefunction-based methods are too expensive to be
applied to the whole system of interest, 
simplified and faster solutions
have to be considered. Such simplications should ideally not alter the
description of 
strong correlation effects. This is where the challenge stands.
Based on the cogent argument
that strong electron correlation is
essentially local~\cite{pulay1983localizability, saebo1993local,
hampel1996local} and that the region of interest is 
one part of a much larger (extended) system,
embedding approaches 
are mainly used in practice~\cite{sun2016quantum}. 
The basic idea is to map the fully-interacting problem onto a so-called 
impurity-interacting one. In the Hubbard model, the impurity corresponds
to an atomic site. 
Among such embedding techniques are the well-established
{\it dynamical mean-field theory}
(DMFT)~\cite{georges1992hubbard,georges1996limitdimension,
kotliar2004strongly,held2007electronic,zgid2011DMFTquantum},
its cluster~
\cite{hettler1998nonlocal,hettler2000dynamical,lichtenstein2000antiferromagnetism,
kotliar2001cellular,maier2005quantum}
and diagrammatic~\cite{rohringer2018diagrammatic} extensions,
as well as combinations of DMFT with either density-functional theory (DFT)
[the so-called
DMFT+DFT approach~\cite{kotliar2006reviewDMFT}] or the Green-function-based GW method
~\cite{sun2002extended,biermann2003first,karlsson2005self,
boehnke2016strong,werner2016dynamical,nilsson2017multitier}. Such
combinations aim at incorporating non-local correlation effects in DMFT.
More recently, the {\it self-energy embedding theory}
(SEET)~\cite{kananenka2015systematically,lan2015communication,zgid2017finite,lan2017testing},
which can be applied to both model and {\it ab initio} Hamiltonians, has
been developed. Let us stress that all the aforementioned embedding
techniques use the frequency-dependent one-particle 
Green function as basic variable.\\

Alternative frequency-independent approaches like the {\it density-matrix embedding theory} 
(DMET)~\cite{knizia2012density,knizia2013density,bulik2014density,zheng2016ground,wouters2016practical,
wouters2016five,rubin2016hybrid} have emerged in recent years. By
construction, standard approximate DMET does not describe 
correlation effects in the environment, thus requiring the treatment 
of more than one impurity site in order to obtain reasonably
accurate results~\cite{knizia2012density}.
More correlation can be incorporated into DMET by using an antisymmetrized geminal power wavefunction~\cite{tsuchimochi2015density},
or, alternatively, by improving the description of the boundary between the fragment and 
the bath~\cite{welborn2016bootstrap}.
Note that Ayral {\it et al.}~\cite{ayral2017dynamical}
succeeded recently in establishing formal connections between DMET, DMFT and the 
{\it rotationally invariant slave bosons} (RISB)
theory~\cite{fresard1992unified,lechermann2007rotationally}.\\ 

This paper deals with another frequency-independent approach, namely {\it site-occupation embedding theory} (SOET)~\cite{fromager2015exact,senjean2017local,senjean2017site}.
While, in conventional
Kohn--Sham (KS) DFT, the fully-interacting problem is mapped onto a
non-interacting one, an auxiliary impurity-interacting system is used in
SOET for extracting the density (i.e. the site occupations in this
context) and, through an appropriate density functional
for the environment (referred to as bath), the total energy. 
From a quantum chemical point of view, SOET is nothing but a 
multi-determinantal extension of KS-DFT for model
Hamiltonians~\cite{chayes1985density,gunnarsson1986density,DFT_ModelHamiltonians}.
In a recent paper~\cite{senjean2017site}, the authors explained how exact
expressions for the double occupation and the per-site energy of the 
uniform Hubbard model can be extracted from SOET. They also proposed
various local density-functional approximations for the bath. The latter work
suffered from two main weaknesses. First of all, the complete
self-consistent formulation of the theory was done only for a single
impurity site, thus preventing a gradual transition from KS-DFT (no
impurity sites) to pure
wavefunction theory (no bath sites). Moreover, none of the proposed DFAs gave
satisfactory results in all correlation and density regimes.\\ 

We explain in this work how these limitations can be overcome. The
paper is
organized as follows. First, an in-principle-exact generalization of SOET to
multiple impurity sites is derived in Sec.~\ref{subsec:soet-M}. The
resulting expressions for the double occupation and the per-site energy
in the uniform case
are given in Sec.~\ref{sec:dblocc_persite}. Existing and newly proposed
DFAs are then discussed in detail in Sec.~\ref{sec:DFAs}. Following the
computational details in Sec.~\ref{sec:comp_details}, results obtained
at half-filling (Sec.~\ref{subsec:half-filling}) and away from
half-filling (Secs.~\ref{subsec:fun_driv_error} and \ref{subsec:SC_results}) are presented
and analyzed. Exact properties of the impurity correlation potential are
discussed in Sec.~\ref{subsec:DD_imp_corr_pot}. Conclusions and
perspectives are finally given in Sec.~\ref{sec:conclu}.

\section{Theory}\label{sec:theory}

\subsection{Site-occupation embedding theory with multiple impurities}\label{subsec:soet-M}

Let us consider the (not necessarily uniform) $L$-site Hubbard Hamiltonian with external potential
$\mathbf{v} \equiv \lbrace v_i \rbrace_{0\leq i\leq L-1}$, 
\begin{eqnarray}\label{eq:Hubb_hamil}
\hat{H}(\mathbf{v}) = \hat{T} + \hat{U} + \hat{V}(\mathbf{v}).
\end{eqnarray}
The hopping operator,   
which is the analog for model Hamiltonians of the kinetic energy
operator,  
reads as follows in second quantization,
\be
\hat{T} = -t \sum_{\langle i,j \rangle}\sum_{\sigma=\uparrow,\downarrow} 
\hat{c}^\dagger_{i
\sigma}\hat{c}_{j\sigma}
,\ee
where $t>0$ is the so-called hopping parameter and $\langle i,j \rangle$ means that the atomic sites $i$ and
$j$ are nearest neighbors. The on-site two-electron
repulsion operator with strength $U$ and the local external potential operator
(which is the analog for model Hamiltonians of the nuclear potential) 
are expressed in terms of the spin-density $\hat{n}_{i\sigma} = 
\hat{c}_{i\sigma}^\dagger \hat{c}_{i\sigma}$ and density $\hat{n}_i = \hat{n}_{i\uparrow} + \hat{n}_{i\downarrow}$
operators as follows,
\be
\hat{U} =    U \sum_i
\hat{n}_{i\uparrow}\hat{n}_{i\downarrow},
\ee
and
\begin{eqnarray}
\hat{V}(\mathbf{v})  & = &  \sum_i v_i \hat{n}_i,
\end{eqnarray}
respectively.\\

The exact ground-state energy $E(\mathbf{v})$ of $\hat{H}(\mathbf{v})$
can be obtained variationally as follows, in complete analogy with
conventional DFT~\cite{hktheo}, 
\begin{eqnarray}\label{ener_min_n}
E(\mathbf{v})= \underset{\mathbf{n}}{\rm min} \left \lbrace F(\mathbf{n}) + 
(\mathbf{v}| \mathbf{n}) \right \rbrace, 
\end{eqnarray}
where $\mathbf{n} \equiv \lbrace n_i \rbrace_{0\leq i\leq L-1}$ is a trial collection of
site occupations
(simply called 
density in the following)
and
$(\mathbf{v}|
\mathbf{n}) = \sum_i v_in_i$.
Within the Levy--Lieb (LL) constrained-search formalism~\cite{levy1979universal}, the
Hohenberg--Kohn functional can be rewritten as follows
in this context,
\begin{eqnarray}\label{eq:LL_full_fun}
F(\mathbf{n}) = \underset{\Psi \rightarrow \mathbf{n}}{\rm min} \left 
\lbrace \langle \Psi \vert \hat{T} + \hat{U} \vert \Psi \rangle \right 
\rbrace
,
 \label{eq:HK}
\end{eqnarray}
where the minimization is restricted to wavefunctions $\Psi$ with density
$\mathbf{n}$. As shown in previous works~\cite{fromager2015exact,senjean2017local,
senjean2017site}, the exact minimizing
density in Eq.~(\ref{ener_min_n}) can be obtained from a fictitious partially-interacting 
system consisting of interacting impurity sites surrounded by
non-interacting ones (the so-called bath sites), thus leading to an in-principle-exact SOET. While
our recent developments focused on the single-impurity version of SOET,
we propose in the following a general formulation of the theory with an  
arbitrary number of impurity
sites. Such a formulation was briefly mentioned in Ref.~\cite{fromager2015exact} for the
purpose of deriving an adiabatic connection formula for the correlation
energy of the bath.\\ 

Let us introduce the analog for
$M$ impurity sites of the LL functional, 
\begin{eqnarray}
F_M^{{\rm imp}}(\mathbf{n}) = \underset{\Psi \rightarrow \mathbf{n}}{\rm min} \left 
\lbrace \left\langle \Psi \middle\vert \hat{T} +
\hat{U}_M
\middle\vert
\Psi \right\rangle \right 
\rbrace
,
 \label{eq:LL_Mimp}
\end{eqnarray}
where
$\hat{U}_M=U\sum^{M-1}_{i=0}\hat{n}_{i\uparrow}\hat{n}_{i\downarrow}$.
Note that, for convenience, the impurity sites have been labelled as
$i=0,1,\ldots,M-1$. If we now introduce the complementary
Hartree-exchange-correlation (Hxc) functional for the bath, 
\begin{eqnarray}\label{eq:Hxc_bath_fun}
\overline{E}_{{\rm Hxc},M}^{\rm 
bath}(\mathbf{n})&=&
F(\mathbf{n})- F_M^{\rm imp}(\mathbf{n})
,
\end{eqnarray}
the ground-state energy expression in Eq.~(\ref{ener_min_n}) can be
rewritten as follows,
\begin{eqnarray}\label{ener_min_psi}
E(\mathbf{v})= \underset{\mathbf{n}}{\rm min} \Bigg\lbrace 
&&\underset{\Psi \rightarrow \mathbf{n}}{\rm min} \left 
\lbrace \left\langle \Psi \middle\vert \hat{T} +
\hat{U}_M
\middle\vert
\Psi \right\rangle \right 
\rbrace
\nonumber\\
&&
+\overline{E}_{{\rm Hxc},M}^{\rm 
bath}(\mathbf{n})
+ 
(\mathbf{v}| \mathbf{n}) \Bigg \rbrace, 
\end{eqnarray}
or, equivalently,
\be
E(\mathbf{v})= \underset{\mathbf{n}}{\rm min} \Bigg\lbrace 
\underset{\Psi \rightarrow \mathbf{n}}{\rm min} 
\Big
\lbrace
&&
\left\langle \Psi \middle\vert \hat{T} +
\hat{U}_M
\middle\vert
\Psi \right\rangle 
\nonumber\\
&&
+\overline{E}_{{\rm Hxc},M}^{\rm 
bath}(\mathbf{n}^\Psi)
+ 
(\mathbf{v}| \mathbf{n}^\Psi)
\Big\rbrace
 \Bigg \rbrace, 
\ee
where
$\mathbf{n}^\Psi\equiv\left\{\langle\Psi\vert\hat{n}_i\vert\Psi\rangle\right\}_{0\leq i\leq L-1}$, 
thus leading to the final variational expression
\be
\label{eq:variational_energy}
E(\mathbf{v})= 
\underset{\Psi}{\rm min} \Big 
\lbrace
&& \left\langle \Psi \middle\vert \hat{T} +
\hat{U}_M
\middle\vert
\Psi \right\rangle 
+\overline{E}_{{\rm Hxc},M}^{\rm 
bath}\left(\mathbf{n}^\Psi\right)
+ 
\left(\mathbf{v}| \mathbf{n}^\Psi\right)
\Big\rbrace
.
\nonumber\\
\ee
The 
minimizing
$M$-impurity-interacting wavefunction
$\Psi_M^{\rm imp}$ in Eq.~(\ref{eq:variational_energy}) reproduces the
exact density profile of the fully-interacting system described by the
Hubbard Hamiltonian in Eq.~(\ref{eq:Hubb_hamil}). From the stationarity
in $\Psi_M^{\rm imp}$ of the energy, we obtain the following 
{\it self-consistent} equation,
\begingroup
\begin{eqnarray}\label{eq:self-consistent-SOET_new}
&&\displaystyle \left( \hat{T} + \hat{U}_{M} + \displaystyle \sum_{i}  
 v^{\rm emb}_{M,i}\,
\hat{n}_i \right) \vert \Psi_M^{\rm imp} \rangle 
 = \mathcal{E}_M^{\rm imp} \vert \Psi_M^{\rm imp} \rangle , 
\end{eqnarray}
\endgroup
where 
\begin{eqnarray}
v^{\rm emb}_{M,i}=v_i+\dfrac{\partial  \overline{E}^{\rm bath}_{{\rm
Hxc},M}(\mathbf{n}^{\Psi_M^{\rm imp}})}{\partial n_i}
\end{eqnarray}
plays 
the role of an embedding
potential for the $M$ impurities. In the particular case of
a uniform half-filled density profile, the embedding potential equals zero in the
bath and $-U/2$ on the impurity sites. This key result, which appears
when applying the hole-particle symmetry transformation to the
impurity-interacting LL functional, is proved in
Appendix~\ref{sec:hole_particle_sym}, thus providing a generalization of
Appendix C in Ref.~\cite{senjean2017site}. Note that
the KS and Schr\"{o}dinger equations are recovered from
Eq.~(\ref{eq:self-consistent-SOET_new}) when $M=0$ and $M=L$ (i.e. the
total number of sites),
respectively. In SOET, $M$ is in the range $0<M<L$, thus leading to a
hybrid formalism where a many-body correlated wavefunction is
embedded into a DFT potential. In practice,
Eq.~(\ref{eq:self-consistent-SOET_new}) can be solved, for example, by
applying an exact diagonalization procedure~\cite{senjean2017local}
(which corresponds to a full configuration interaction) for small
rings, or by using the more advanced {\it density matrix renormalization
group} (DMRG) method which allows for the
description of larger systems~\cite{senjean2017site}.\\

Let us now return to the expression in Eq.~(\ref{eq:Hxc_bath_fun}) of the
complementary Hxc energy for the
bath. By using the KS decompositions of the fully-interacting and
$M$-impurity-interacting LL functionals,  
\begin{eqnarray}
F(\mathbf{n}) = T_{\rm s}(\mathbf{n}) + E_{\rm Hxc}
(\mathbf{n})
\end{eqnarray}
and
\begin{eqnarray}\label{eq:KS_decomp_Mimp}
F_M^{\rm imp}(\mathbf{n}) = T_{\rm s}(\mathbf{n}) + E^{\rm imp}_{{\rm
Hxc},M}
(\mathbf{n}), 
\end{eqnarray}
respectively, 
where the non-interacting kinetic energy functional reads as follows in
this context,
\begin{eqnarray}
T_{\rm s}(\mathbf{n}) = \underset{\Psi \rightarrow 
\mathbf{n}}{\rm min}  \{ \langle \Psi \vert \hat{T}\vert \Psi 
\rangle \},
\end{eqnarray}
we obtain 
\be\overline{E}_{{\rm Hxc},M}^{\rm 
bath}(\mathbf{n})=E_{\rm Hxc}
(\mathbf{n})-E^{\rm imp}_{{\rm
Hxc},M}
(\mathbf{n}).
\ee  
If we now separate the Hxc energies into Hx (i.e. mean-field) and
correlation contributions,
\be
E_{\rm Hxc}(\mathbf{n})&=&\dfrac{U}{4}\sum_{i} n_i^2 +
E_{\rm c}(\mathbf{n}),
\ee
and
\be\label{eq:Hx_plus_c_imp}
E^{\rm imp}_{{\rm
Hxc},M}
(\mathbf{n})&=&\dfrac{U}{4}\sum^{M-1}_{i=0} n_i^2+E^{\rm imp}_{{\rm
c},M}
(\mathbf{n}),
\ee
we obtain the final expression
\begin{eqnarray}\label{eq:bathHxc_fun_M-imp}
\overline{E}_{{\rm Hxc},M}^{\rm
bath}(\mathbf{n})= \dfrac{U}{4}\sum_{i \geqslant M} n_i^2 +
\overline{E}_{{\rm c},M}^{{\rm bath}}(\mathbf{n}),
\end{eqnarray}
where 
\begin{eqnarray}
\label{eq:Ecbath_expression}
\overline{E}_{{\rm c},M}^{\rm bath}(\mathbf{n})=E_{\rm c}(\mathbf{n}) -
E_{{\rm 
c},M}^{\rm imp} (\mathbf{n}).
\end{eqnarray}
While local density approximations (LDA) based, for example, on the Bethe
ansatz (BALDA) are available for $E_{\rm c}(\mathbf{n})$ in the
literature~\cite{lima2002density,lima2003density, capelle2003density}, no DFA has been developed so
far for modeling the correlation energy of multiple impurity sites.
Existing approximations for a
single impurity are reviewed in Sec.~\ref{sec:dfa_single_imp}. Newly
proposed DFAs will be introduced in Secs.~\ref{sec:2L-BALDA} and~\ref{sec:dfa_multiple_imp}.

\subsection{Exact double occupation and per-site energy expressions in the uniform case}\label{sec:dblocc_persite}

In this section we derive exact SOET expressions for the per-site energy
and double occupancy in the particular case of the uniform Hubbard
system ($\mathbf{v} =\mathbf{0}$) for which the LDA decomposition of the
full correlation energy in terms of per-site contributions,
\be\label{eq:lda_full_corr_ener}
E_{\rm c}(\mathbf{n}) = \sum_i e_{\rm c}(n_i),
\ee
is exact. Thus we extend to multiple
impurities Eqs.~(21) and (23) of Ref.~\cite{senjean2017site}.\\
 
For that
purpose, let us introduce the following {\it per-site} analog of
Eq.~(\ref{eq:Ecbath_expression}),
\be\label{eq:ecbath_per_site_M}
\overline{e}_{{\rm c},M}^{\rm
bath}(\mathbf{n})=\dfrac{1}{M}\left[\left(\sum^{M-1}_{i=0}e_c(n_i)\right)-E_{{\rm 
c},M}^{\rm imp} (\mathbf{n})\right]
,\ee
which, for a uniform density profile $\underline{n}=(n,n,\ldots,n)$,
gives   
\be\label{eq:ecbath_per_site_M_uniform}
\overline{e}_{{\rm c},M}^{\rm
bath}(\underline{n})=e_c(n)-\dfrac{
E_{{\rm 
c},M}^{\rm imp} (\underline{n})
}{M}
.\ee
Note that, when combining Eqs.~(\ref{eq:lda_full_corr_ener})
and~(\ref{eq:ecbath_per_site_M}) with Eq.~(\ref{eq:Ecbath_expression}),
we obtain the following expression for the bath correlation energy,
\be\label{eq:Ecbath=sum_ec+Mecbath}
\overline{E}_{{\rm c},M}^{\rm bath}(\mathbf{n})=\sum_{i \geqslant M}e_{\rm
c}(n_i)+M\overline{e}_{{\rm c},M}^{\rm
bath}(\mathbf{n}).
\ee 
By inserting the decomposition in Eq.~(\ref{eq:ecbath_per_site_M_uniform}) into the exact double
site-occupation expression~\cite{DFT_ModelHamiltonians} (we denote
$E=E({\mathbf{v}=\mathbf{0}})$ for simplicity),
\begin{eqnarray}
d=\langle\hat{n}_{i\uparrow}\hat{n}_{i\downarrow}\rangle=\dfrac{1}{L}\dfrac{\partial E}{\partial U}=\dfrac{n^2}{4}+\dfrac{\partial e_{\rm
c}(n)}{\partial U}, \label{eq:dblocc_KS}
\end{eqnarray}   
where $n=N/L$ and $N$ is the total
number of electrons, it comes from Eq.~(\ref{eq:Hx_plus_c_imp}),
\be\label{eq:2ble_occ_M-soet}
d=\dfrac{1}{M}\dfrac{\partial E_{{\rm
Hxc},M}^{\rm imp} (\underline{n})}{\partial U}
+\dfrac{\partial \overline{e}_{{\rm c},M}^{\rm
bath}(\underline{n})}{\partial U}
.\ee
By using the fact that, in the particular (uniform) case considered here,
$E=F(\underline{n})$ and, according to the Hellmann--Feynman theorem
(see Eq.~(\ref{eq:variational_energy})),
\be
\dfrac{\partial E}{\partial
U}=\sum^{M-1}_{i=0}
\langle\hat{n}_{i\uparrow}\hat{n}_{i\downarrow}\rangle_{\Psi_M^{\rm imp}}
+
\dfrac{\partial \overline{E}_{{\rm
Hxc},M}^{\rm bath} (\underline{n})}{\partial U}
,\ee
where 
\be\label{eq:exact_density_imp_system}
\left\langle\Psi_M^{\rm
imp}\middle\vert\hat{n}_i\middle\vert\Psi_M^{\rm imp}\right\rangle=n
\ee
for any site $i$, it comes from the separations in
Eqs.~(\ref{eq:Hxc_bath_fun}) and (\ref{eq:KS_decomp_Mimp}) that  
\be\label{eq:sum_2ble_occ_fun}
\dfrac{\partial }{\partial
U}\left[E-
\overline{E}_{{\rm
Hxc},M}^{\rm bath} (\underline{n})
\right]
&=&\sum^{M-1}_{i=0}
\langle\hat{n}_{i\uparrow}\hat{n}_{i\downarrow}\rangle_{\Psi_M^{\rm imp}}
\nonumber\\
&=&
\dfrac{\partial F_M^{\rm imp}(\underline{n})}{\partial U}
\nonumber\\
&=&
\dfrac{\partial E_{{\rm
Hxc},M}^{\rm imp} (\underline{n})}{\partial U}
.\ee
Finally, combining Eqs.~(\ref{eq:2ble_occ_M-soet}), (\ref{eq:exact_density_imp_system}) and
(\ref{eq:sum_2ble_occ_fun}) leads to the following {\it exact} expression for
the double occupation in SOET with multiple impurities,
\be\label{eq:dblocc_SOET}
d=\dfrac{1}{M}
\sum^{M-1}_{i=0}
\langle\hat{n}_{i\uparrow}\hat{n}_{i\downarrow}\rangle_{\Psi_M^{\rm imp}}
+\dfrac{\partial \overline{e}_{{\rm c},M}^{\rm
bath}({\bf n}^{\Psi_M^{\rm imp}})}{\partial U}
.\ee
The expression derived in Ref.~\cite{senjean2017site} in the particular case of a
single impurity site is recovered from Eq.~(\ref{eq:dblocc_SOET}) when $M=1$. 
Note also that the double occupations of the impurity sites are in
principle {\it not} equal to each other in the fictitious $M$-impurity-interacting system, simply because translation symmetry is
broken, as readily seen from Eq.~(\ref{eq:self-consistent-SOET_new}), even
though the embedding
potential restores uniformity in the density profile.\\

Turning to the per-site
energy~\cite{lima2003density}, 
\begin{eqnarray}\label{eq:per-site-energy_KS}
e=E/L=t_{\rm s}(n)+\dfrac{U}{4}n^2+e_{\rm c}(n),
\end{eqnarray}   
where $t_{\rm s}(n)$ is the per-site non-interacting kinetic energy
functional,
we can insert Eq.~(\ref{eq:ecbath_per_site_M_uniform}) into
Eq.~(\ref{eq:per-site-energy_KS}) and use Eqs.~(\ref{eq:KS_decomp_Mimp})
and (\ref{eq:Hx_plus_c_imp}), thus leading to 
\be\label{eq:per_site_ener_fun_M-soet}
e=t_{\rm s}(n)+\dfrac{1}{M}\left(F_M^{\rm imp}(\underline{n})-T_{\rm
s}(\underline{n})\right)+\overline{e}_{{\rm c},M}^{\rm
bath}(\underline{n}).
\ee
Moreover, applying once more the Hellmann--Feynman theorem to the variational energy
expression in Eq.~(\ref{eq:variational_energy}) gives
\be
t\dfrac{\partial E}{\partial t}=\left\langle\Psi_M^{\rm
imp}\middle\vert\hat{T}\middle\vert\Psi_M^{\rm imp}\right\rangle
+t\dfrac{\partial \overline{E}_{{\rm Hxc},M}^{\rm
bath}(\underline{n})}{\partial t},\ee
or,
equivalently,
\be\label{eq:Timp_uniform}
\left\langle\Psi_M^{\rm
imp}\middle\vert\hat{T}\middle\vert\Psi_M^{\rm
imp}\right\rangle=t\dfrac{\partial F_M^{\rm
imp}(\underline{n})}{\partial t},
\ee
which, for $U=0$, leads to
\be
T_{\rm s}(\underline{n})=t\dfrac{\partial T_{\rm s}(\underline{n})}{\partial t}
.\ee
As a result, the second term in the right-hand side of Eq.~(\ref{eq:per_site_ener_fun_M-soet}) can
be simplified as follows,
\be
&&F_M^{\rm imp}(\underline{n})-T_{\rm
s}(\underline{n})=\left\langle\Psi_M^{\rm
imp}\middle\vert\hat{T}\middle\vert\Psi_M^{\rm
imp}\right\rangle
-T_{\rm
s}(\underline{n})
\nonumber\\
&&
+U\sum^{M-1}_{i=0}
\langle\hat{n}_{i\uparrow}\hat{n}_{i\downarrow}\rangle_{\Psi_M^{\rm imp}}
\nonumber\\
&&
=t\dfrac{\partial {E}_{{\rm c},M}^{\rm
imp}(\underline{n})}{\partial t}
+U\sum^{M-1}_{i=0}
\langle\hat{n}_{i\uparrow}\hat{n}_{i\downarrow}\rangle_{\Psi_M^{\rm imp}}
,\ee
thus leading, according to Eq.~(\ref{eq:ecbath_per_site_M_uniform}), to
the final {\it exact} per-site energy
expression,  
\be\label{eq:per_site_ener_M-soet_final}
e&=&t_{\rm s}(n)+\dfrac{U}{M}\sum^{M-1}_{i=0}
\langle\hat{n}_{i\uparrow}\hat{n}_{i\downarrow}\rangle_{\Psi_M^{\rm imp}}
+t\dfrac{\partial e_{\rm c}(n)}{\partial t}
\nonumber\\
&&-t\dfrac{\partial \overline{e}_{{\rm c},M}^{\rm
bath}(\underline{n})}{\partial t}
+\overline{e}_{{\rm c},M}^{\rm
bath}(\underline{n}),
\ee
or, equivalently,
\be\label{eq:per_site_ener_M-soet_final3}
e&=&
\dfrac{1}{M}\left.
\sum^{M-1}_{i=0}
\left[
t_{\rm s}(n_i^\Psi)+t\dfrac{\partial e_{\rm c}(n_i^\Psi)}{\partial
t}
+{U}
\langle\hat{n}_{i\uparrow}\hat{n}_{i\downarrow}\rangle_{\Psi}
\right]
\right|_{\Psi=\Psi_M^{\rm imp}}
\nonumber\\
&&
+\left.\left[\overline{e}_{{\rm c},M}^{\rm
bath}({\bf n}^\Psi)
-t\dfrac{\partial \overline{e}_{{\rm c},M}^{\rm
bath}({\bf n}^\Psi)}{\partial t}
\right]
\right|_{\Psi={\Psi_M^{\rm imp}}}
.
\ee
%
%
Note that the expression in Eq.~(\ref{eq:per_site_ener_M-soet_final3}),
which is a generalization for multiple impurities of the energy
expression derived in Ref.~\cite{senjean2017site}, is convenient for practical
(approximate) SOET calculations where the density profile calculated
self-consistently might deviate significantly from uniformity~\cite{senjean2017site}.
Finally, as shown in Appendix~\ref{appendix:fun_rel}, since the exact
per-site bath correlation functional fulfills the fundamental relation,
\be\label{eq:fun_rel_ecbath-M-imp}
\overline{e}_{{\rm c},M}^{\rm
bath}({\bf n})&=&
t\dfrac{\partial \overline{e}_{{\rm c},M}^{\rm
bath}({\bf n})}{\partial
t}
+{U}
\dfrac{\partial \overline{e}_{{\rm c},M}^{\rm
bath}({\bf n})}{\partial
U},
\ee
Eq.~(\ref{eq:per_site_ener_M-soet_final3}) can be further
simplified as follows,
\be\label{eq:per_site_ener_M-soet_final4}
e&=&
\dfrac{1}{M}\left.
\sum^{M-1}_{i=0}
\left[
t_{\rm s}(n_i^\Psi)+t\dfrac{\partial e_{\rm c}(n_i^\Psi)}{\partial
t}
+{U}
\langle\hat{n}_{i\uparrow}\hat{n}_{i\downarrow}\rangle_{\Psi}
\right]
\right|_{\Psi=\Psi_M^{\rm imp}}
\nonumber\\
&&
+\left.U
\dfrac{\partial \overline{e}_{{\rm c},M}^{\rm
bath}({\bf n}^\Psi)}{\partial U}
\right|_{\Psi={\Psi_M^{\rm imp}}}
.
\ee
Let us stress that any {\it approximate} density functional of the form
$t\,\times\,\mathcal{G}(U/t,{\bf n})$ 
fulfills the exact condition in
Eq.~(\ref{eq:fun_rel_ecbath-M-imp}). This is the case for all the DFAs
considered in this work. As a result, switching from
Eq.~(\ref{eq:per_site_ener_M-soet_final3}) to
Eq.~(\ref{eq:per_site_ener_M-soet_final4}) brings no additional errors
when approximate functionals are used.

\section{Local density-functional approximations in SOET}\label{sec:DFAs}

In order to perform practical SOET calculations and compute, for
example, per-site energies, we need
DFAs, not only for the per-site correlation
energy $e_{\rm
c}({{n}})$, like in conventional KS-DFT, but also for the
per-site complementary bath correlation energy $\overline{e}^{\rm bath}_{{\rm
c},M}(\mathbf{n})$ or, equivalently, for the impurity correlation energy $E_{{\rm
c},M}^{\rm imp} (\mathbf{n})$ [see Eq.~(\ref{eq:ecbath_per_site_M})].  
Existing approximations to the latter functionals are discussed in
Secs.~\ref{subsubsec:balda} and ~\ref{sec:dfa_single_imp}, respectively. 
A new functional is proposed in Sec.~\ref{sec:2L-BALDA} and a simple
multiple-impurity DFA  
is introduced in Sec.~\ref{sec:dfa_multiple_imp}.
Let us stress that, in all the DFAs considered in this work, we make the
approximation that the impurity
correlation functional does not depend on the occupations in the bath,
\be\label{eq:ilda_M-imp}
E_{{\rm 
c},M}^{\rm imp} (\mathbf{n})\rightarrow E_{{\rm 
c},M}^{\rm imp} (\mathbf{n}_{\rm imp}),
\ee
or, equivalently,
\begin{eqnarray}\label{eq:ILDA}
\overline{e}_{{\rm c},M}^{\rm bath}(\mathbf{n}) \rightarrow
\overline{e}_{{\rm c},M}^{\rm bath}(\mathbf{n}_{\rm imp}),
\end{eqnarray}
where $\mathbf{n}_{\rm imp}\equiv(n_0,n_1,\ldots,n_{M-1})$ is the
collection of densities on the 
impurity sites. The implications of such an approximation are discussed
in detail in Ref.~\cite{senjean2017local}. 

\subsection{Bethe ansatz LDA for $e_{\rm c}({{n}})$}\label{subsubsec:balda}

The BALDA approximation~\cite{lima2002density,lima2003density, capelle2003density} to the full per-site correlation energy functional $e_{\rm c}({{n}})$ is exact
for $U = 0$, $U \rightarrow +\infty$,
and all $U$ values at half-filling ($n=1$). It reads as follows,
\begin{eqnarray}\label{eq:ecBALDA}
e_{\rm c}^{\rm BALDA}(U,t,n)&  =  & e^{\rm BALDA}(U,t,n)
\nonumber\\
&& -  e^{\rm BALDA}(U=0,t,n)
- \dfrac{U}{4}n^2,
\end{eqnarray}
where the BALDA density-functional energy equals
\begin{eqnarray}\label{eq:eBALDA_nleq1}
e^{\rm BALDA}(U,t,n \leqslant1) = \dfrac{-2t \beta (U/t)}{\pi} \sin \left( \dfrac{\pi n}{\beta (U/t)} \right),
\end{eqnarray}
and
\begin{eqnarray}\label{eq:eBALDA_ngeq1}
e^{\rm BALDA}(U,t,n \geqslant 1) & = & e^{\rm BALDA}(U,t,2 - n) 
\nonumber\\
&&+ U(n - 1). 
\end{eqnarray}
The $U/t$-dependent function $\beta (U/t)$ is determined by solving
\begin{eqnarray}\label{eq:beta}
 \dfrac{-2 \beta (U/t)}{\pi} \sin \left( \dfrac{\pi}{\beta (U/t)} \right)
 =  -4 \int_0^\infty {\rm d}x \dfrac{J_0(x)J_1(x)}{x(1 + \exp(Ux/2t))}, \nonumber \\
\end{eqnarray}
where $J_0$ and $J_1$ are zero- and first-order Bessel functions.
When $U=0$, the BALDA energy reduces to the (one-dimensional)
non-interacting kinetic energy functional
\be\label{eq:ts_fun}
t_{\rm s}(n)=-4t\sin(\pi n/2)/\pi,
\ee
which is exact in the thermodynamic limit ($L\rightarrow +\infty$).

\subsection{Review of existing DFAs for a single impurity}\label{sec:dfa_single_imp}

\subsubsection{Impurity-BALDA}

The impurity-BALDA approximation (iBALDA), which was originally formulated in
Ref.~\cite{senjean2017site} for a single impurity site,
consists in modeling the correlation energy of the impurity-interacting
system with BALDA:
\begin{eqnarray}\label{eq:iBALDA_Ecimp}
E_{{\rm c},M=1}^{\rm imp}(\mathbf{n}) \xrightarrow{\rm iBALDA} e_{\rm c}^{\rm BALDA}(n_0).
\end{eqnarray}
In other
words, 
the iBALDA
neglects the contribution of the bath to the total per-site correlation
energy [see Eq.~(\ref{eq:ecbath_per_site_M}) with $M$=1], 
\begin{eqnarray}\label{eq:ibalda_approx}
\overline{e}_{{\rm c},M=1}^{\rm bath}(\mathbf{n}) \xrightarrow{\text{iBALDA}} 0.
\end{eqnarray}

\subsubsection{DFA based on the single-impurity Anderson model.}

As shown in Ref.~\cite{senjean2017site}, a simple single-impurity
correlation functional can be designed from the following perturbation expansion
in $U/\Gamma$ of the symmetric {\it single-impurity Anderson model}~(SIAM) correlation energy~\cite{yamada1975perturbation},
\begin{eqnarray}\label{eq:anderson_secondorder}
E_{{\rm c}, U/\Gamma \rightarrow 0}^{\rm SIAM}(U,\Gamma) = 
\dfrac{U^2}{\pi \Gamma} \left[ - 0.0369 + 0.0008\left(\dfrac{U}{\pi 
\Gamma}\right)^2\right], \nonumber \\
\end{eqnarray}
where $\Gamma$ is the so-called impurity level
width parameter of the SIAM. A density functional is obtained from
Eq.~(\ref{eq:anderson_secondorder}) by introducing the following
$t$-dependent density-functional impurity level
width,
\begin{eqnarray}\label{eq:imp_level_width}
\Gamma(t,n) = t\left( \dfrac{1 + \cos (\pi n /2)}{\sin (\pi n/2)}\right).
\end{eqnarray}
A rationale for this choice is given in Ref.~\cite{senjean2017site}.
Combining the resulting impurity correlation functional with BALDA gives the so-called SIAM-BALDA
approximation~\cite{senjean2017site}. In summary, within SIAM-BALDA, we
make the following approximations,
\begin{eqnarray}\label{eq:SIAM-BALDA}
E_{{\rm c},M=1}^{\rm imp}(\mathbf{n}) \xrightarrow{\text{SIAM-BALDA}} 
E_{{\rm c}, U/\Gamma \rightarrow 0}^{\rm SIAM}\Big(U,\Gamma(t,n_0)\Big), 
\end{eqnarray}
and
\be
\overline{e}_{{\rm c},M=1}^{\rm
bath}(\mathbf{n})&\xrightarrow{\text{SIAM-BALDA}}&e_{\rm c}^{\rm
BALDA}(n_0)
\nonumber\\
&&-E_{{\rm c}, U/\Gamma \rightarrow 0}^{\rm
SIAM}\Big(U,\Gamma(t,n_0)\Big).
\nonumber\\
\ee

\subsection{Combined two-level/Bethe ansatz LDA functional}\label{sec:2L-BALDA}

As shown in Ref.~\cite{senjean2017local}, in the particular case of the
two-level (2L) Hubbard model (also referred to as the Hubbard dimer)
with two electrons, the full density-functional correlation
energy $E_{\rm c}^{\rm 2L}(U,n)$ is connected to the impurity one by a
simple scaling relation,
\begin{eqnarray}\label{eq:dimer_corr}
E^{\rm imp,2L}_{{\rm c}}(U,n_0) = E_{\rm c}^{\rm 2L}(U/2,n_0),
\end{eqnarray}
where $n_0$ is the occupation of the impurity site. In this case, the
bath reduces to a single site with occupation $n_1=2-n_0$. Combining
Eq.~(\ref{eq:dimer_corr}) with BALDA gives us a new single-impurity DFA that will be
referred to as 2L-BALDA in the following. In summary, within 2L-BALDA,
we make the following approximations, 
\begin{eqnarray}\label{eq:2LBALDA}
E_{{\rm c},M=1}^{\rm imp}(\mathbf{n})  \xrightarrow{\rm 2L-BALDA} E^{\rm
2L}_{{\rm c}}(U/2,n_0), 
\end{eqnarray}
and
\be
\overline{e}_{{\rm c},M=1}^{\rm
bath}(\mathbf{n})&\xrightarrow{\text{2L-BALDA}}&e_{\rm c}^{\rm
BALDA}(n_0)
\nonumber\\
&&-
E^{\rm
2L}_{{\rm c}}(U/2,n_0).
\ee
In our calculations, the accurate parameterization of Carrascal
{\etal}~\cite{carrascal2015hubbard,carrascal2016corrigendum} has been
used for $E_{\rm c}^{\rm 2L}(U,n)$. 

\subsection{DFA for multiple impurity sites}\label{sec:dfa_multiple_imp}

As pointed out in Sec.~\ref{subsec:soet-M}, in SOET, one can gradually move from
KS-DFT to pure wavefunction theory by increasing the number $M$ of
impurities from 0 to the number $L$ of sites. Let us, for convenience,
introduce the following notation,
\be
\overline{e}_{{\rm c},M}^{\rm bath}(\mathbf{n})=
\overline{\varepsilon}_{{\rm c},M/L}^{\rm bath}(\mathbf{n}),
\ee
or, equivalently,
\be
\overline{\varepsilon}_{{\rm c},\mathcal{M}}^{\rm
bath}(\mathbf{n})=\overline{e}_{{\rm c},L\mathcal{M}}^{\rm
bath}(\mathbf{n}),
\ee
where $\mathcal{M}=M/L$ is the proportion of impurity sites in the
partially-interacting system. In the thermodynamic limit, $\mathcal{M}$ becomes a continuous
variable and, if the number of impurity sites is large enough, the
following Taylor expansion can be used,
\be\label{PT_in_nbr_imp}
\overline{\varepsilon}_{{\rm c},\mathcal{M}}^{\rm
bath}(\mathbf{n})&=&\overline{\varepsilon}_{{\rm c},\mathcal{M}=1}^{\rm
bath}(\mathbf{n})+\left.\dfrac{\partial \overline{\varepsilon}_{{\rm c},\mathcal{M}}^{\rm
bath}(\mathbf{n})}{\partial
\mathcal{M}}\right|_{\mathcal{M}=1}\times(\mathcal{M}-1)
\nonumber\\
&&+\mathcal{O}\Big(\left(\mathcal{M}-1\right)^2\Big).
\ee
As readily seen from Eqs.~(\ref{eq:LL_full_fun}), (\ref{eq:LL_Mimp}),
(\ref{eq:Hxc_bath_fun}), (\ref{eq:bathHxc_fun_M-imp}), and
(\ref{eq:Ecbath=sum_ec+Mecbath}), when $\mathcal{M}=1$ or, equivalently,
$M=L$, we have
\be
\overline{\varepsilon}_{{\rm c},\mathcal{M}=1}^{\rm
bath}(\mathbf{n})=0.
\ee
If, for simplicity, we keep the zeroth-order term only in
Eq.~(\ref{PT_in_nbr_imp}), we obtain a generalization of iBALDA for $M$
impurities, which is denoted as iBALDA($M$) in the following. The
exploration of first- and higher-order corrections in
Eq.~(\ref{PT_in_nbr_imp}) is left for future
work. In summary, within iBALDA($M$), we make the following
approximations,
\be
\overline{e}_{{\rm c},M}^{\rm bath}(\mathbf{n}) \xrightarrow{{\rm
iBALDA}(M)}0,
\ee
and
\be
E_{{\rm c},M}^{\rm imp}(\mathbf{n})\xrightarrow{{\rm
iBALDA}(M)}
\sum_{i=0}^{M-1} e_{\rm c}^{\rm BALDA}(n_i).
\ee

\section{Computational details}\label{sec:comp_details}

The various DFAs discussed in Sec.~\ref{sec:DFAs} and
summarized in Table~\ref{tab} have been applied to
the $L$-site uniform one-dimensional
Hubbard model with an even number $N$ of electrons and $L=32$.
Periodic ($\hat{a}_{L\sigma}=
\hat{a}_{0\sigma}$) and 
antiperiodic
($\hat{a}_{L\sigma}
=-\hat{a}_{0\sigma}$) boundary conditions have been used when   
$(N/2)~{\rm mod}~2 = 1$ [i.e. $N/2$ is an odd number]
and $(N/2)~{\rm mod}~2 = 0$ [i.e. $N/2$ is an even 
number], respectively.
In all the SOET calculations [note that, in the following, they will be referred to by the name of the
DFA that is employed (see the first column of Table~\ref{tab})], the 
DMRG method~\cite{white1992density,white1993density,verstraete2008matrix,schollwock2011density,naokicode} 
has been 
used for solving (self-consistently or with the exact uniform density)
the many-body Eq.~(\ref{eq:self-consistent-SOET_new}).
The maximum number of renormalized states (or virtual bond
dimension) was set to $m = 500$. 
Standard DMRG calculations (simply referred to as DMRG in the following) have also been performed on the conventional
uniform Hubbard system for comparison. 
For analysis purposes, exact correlation energies and their derivatives in $t$ and $U$
have been computed in a smaller 8-site ring. Technical details are given
in Appendix~\ref{appendix:Lieb_max}.
The performance of the various DFAs has been evaluated by computing
double occupations and per-site energies according to
Eqs.~(\ref{eq:dblocc_SOET}) and (\ref{eq:per_site_ener_M-soet_final4}),
respectively. This required the implementation of various DFA derivatives.
Details about the derivations are given in  
Appendices~\ref{sec:derivatives_BALDA},~\ref{sec:derivatives_SIAM-BALDA},
and \ref{sec:derivatives_2L-BALDA}.
Note that the hopping parameter has been set to $t = 1$ in all the
calculations.

\begin{table*}
\begin{center}
\begin{tabularx}{1\textwidth}{c *{2}{Y}}
\toprule
\hline\hline
 \multicolumn{1}{c}{SOET method}
 & \multicolumn{1}{c}{DFA used for $\overline{E}_{{\rm Hxc},M}^{\rm bath}(\mathbf{n})$} 
 & \multicolumn{1}{c}{Correlation functionals} \\ \\
\hline
 \\
iBALDA($M$) & $\displaystyle \sum_{i=M}^{L-1} \left[\dfrac{U}{4}n_i^2 +
e_{\rm c}^{\rm BALDA}(n_i)\right]$
& Eqs.~(\ref{eq:ecBALDA})--(\ref{eq:beta})\\ \\
SIAM-BALDA & $\displaystyle \sum^{L-1}_{i=0} \left[\dfrac{U}{4}n_i^2 + e_{\rm
c}^{\rm BALDA}(n_i) \right] 
{-\dfrac{U}{4}n_0^2-E^{\rm SIAM}_{{\rm c},U/\Gamma\rightarrow
0}\left(U,\Gamma(t,n_0)\right)}$
&Eqs.~(\ref{eq:ecBALDA})--(\ref{eq:beta}),
(\ref{eq:anderson_secondorder}) and (\ref{eq:imp_level_width})\\ \\
2L-BALDA & $\displaystyle \sum^{L-1}_{i=0} \left[\dfrac{U}{4}n_i^2 +
e_{\rm c}^{\rm BALDA}(n_i) \right] 
-\dfrac{U}{4}n_0^2-E_{\rm c}^{{\rm 2L}}(U/2,n_0)$
&Eqs.~(\ref{eq:ecBALDA})--(\ref{eq:beta}) and (\ref{eq:Ec_burke}) 
\\ \\
\hline
\hline
\end{tabularx}
\caption{Summary of the single- and multiple-impurity Hxc DFAs used in
this work for the
bath. See Sec.~\ref{sec:DFAs} for further details.}
\label{tab}
\end{center}
\end{table*}

\begin{figure}
\resizebox{0.49\textwidth}{!}{
\includegraphics[scale=1]{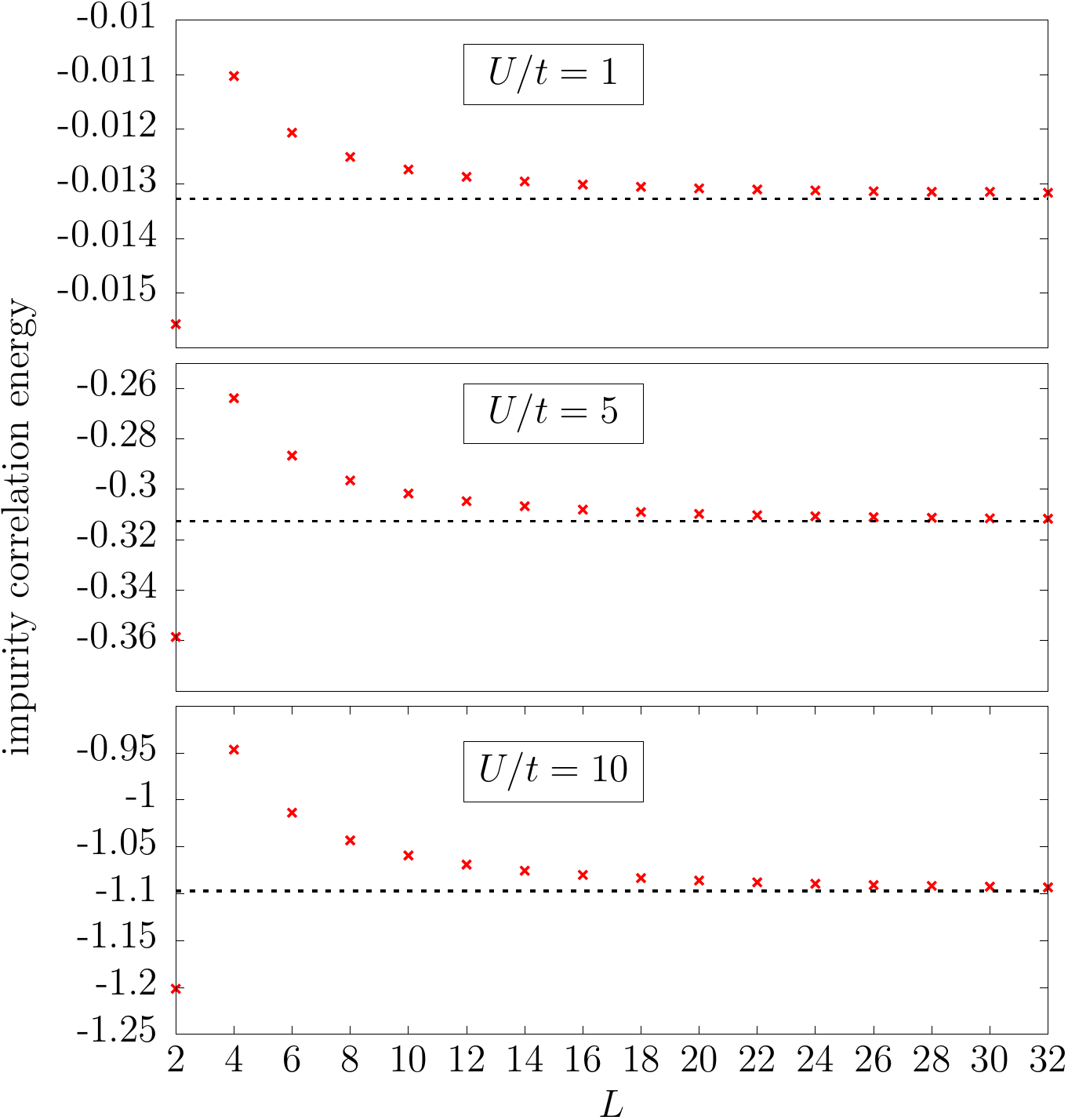}
}
\caption{Accurate single-impurity ($M=1$) correlation density-functional energies
computed at the DMRG level in the half-filled case. 
Results obtained for $L = 64$ are shown in black dashed lines. See text
for further details.}
\label{fig:Ecimp_fctL}
\end{figure}

\section{Results and discussion}\label{sec:results}


\subsection{Half-filled case}\label{subsec:half-filling}


In this work, SOET is applied to a relatively small 32-site ring.
Nevertheless, as illustrated in the following, the number of sites is
large enough so that there are no substantial
finite-size errors in the calculation of density-functional energies. This can be easily seen in the half-filled case
($n=1$) where the exact embedding
potential in Eq.~(\ref{eq:self-consistent-SOET_new}) equals zero in the
bath and $-U/2$ on the impurity sites (see Appendix~\ref{sec:hole_particle_sym}).
The resulting impurity correlation density-functional energy [see Appendix~\ref{appendix:Lieb_max} for
further details] calculated at the DMRG level is
shown in Fig.~{\ref{fig:Ecimp_fctL}. The convergence towards the
thermodynamic limit ($L \rightarrow + \infty$) is relatively fast. 
Note that the results obtained for $L=32$ and $L=64$ are almost
undistinguishable.\\

Let us now focus on the calculation of double occupations which has
been implemented according to Eq.~(\ref{eq:dblocc_SOET}) for various
approximate functionals. Results are shown in Fig.~\ref{fig:dblocc}.
\begin{figure}
\resizebox{0.49\textwidth}{!}{
\includegraphics[scale=1]{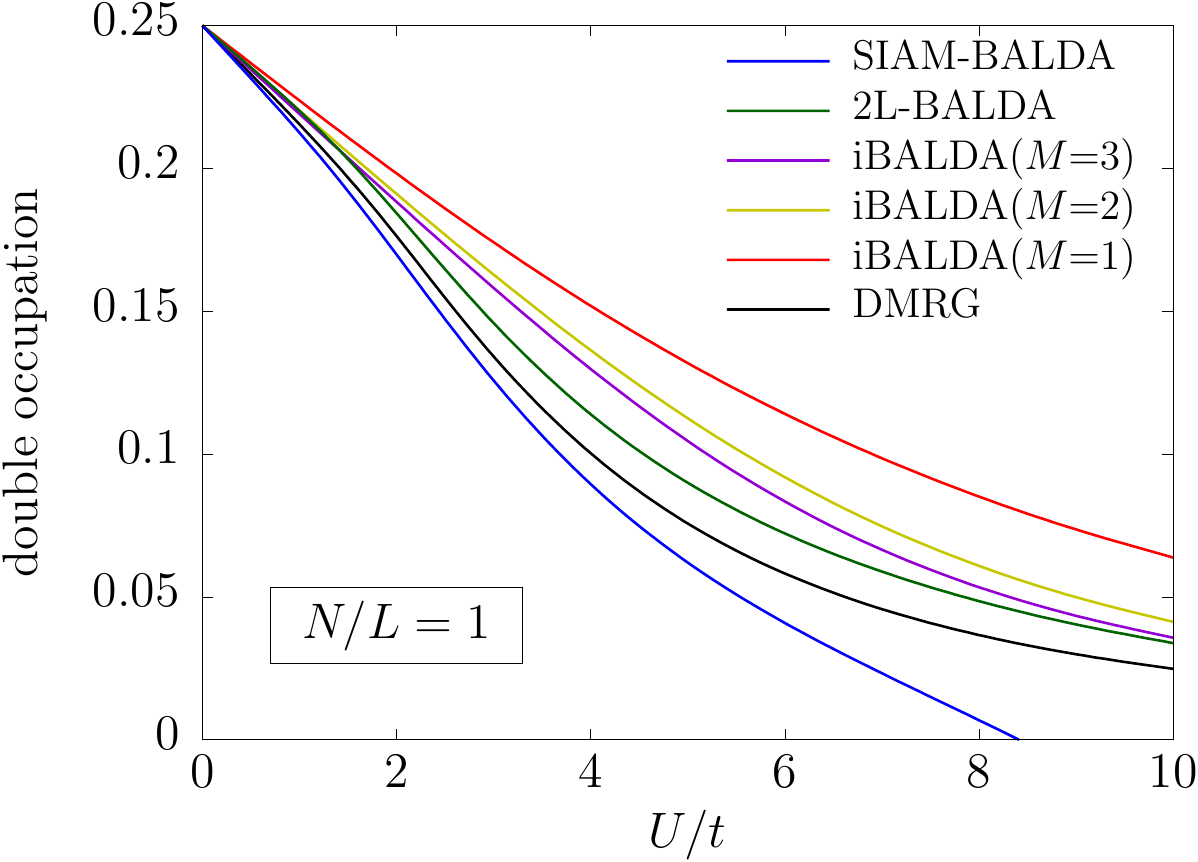}
}
\caption{
Double occupations obtained from half-filled SOET calculations for various single-
and multiple-impurity DFAs. Both standard reference DMRG and iBALDA($M$=1)
results are taken from Ref.~\cite{senjean2017site}. See text for further
details.
}
\label{fig:dblocc}
\end{figure}
While, in the weakly correlated regime and up to $U/t=5$, SIAM-BALDA
gives the best results, it dramatically fails in stronger correlation
regimes, as expected~\cite{senjean2017site}. In the particular
strongly correlated half-filled case, SIAM-BALDA can be improved
by interpolating between the weakly and strongly
correlated regimes of the SIAM~\cite{senjean2017site}. Unfortunately,
generalizing such an interpolation away from
half-filling is not straightforward~\cite{senjean2017site}. 
On the other hand, 2L-BALDA
performs relatively well for all the values of $U/t$. Turning to the multiple-impurity iBALDA approximation, the
accuracy increases with the number of impurities, as expected, but the
convergence towards DMRG is slow. Switching from two to three
impurities slightly improves on the result, which is still less accurate
than the (single-impurity) 2L-BALDA one. Interestingly, the same pattern
is observed in DMET when the matching criterion involves 
the impurity site occupation only [see the noninteracting (NI) bath
formulation in Ref.~\cite{bulik2014density} and the Fig.~2 therein].
While our results would be improved by designing better $M$-dependent
DFAs than iBALDA($M$), the performance of multiple-impurity DMET
is increased when matching not only diagonal but also
non-diagonal density matrix elements~\cite{knizia2012density,bulik2014density}.
Further connections between SOET and DMET are currently investigated and will
be presented in a separate work.\\
\begin{figure}
\resizebox{0.49\textwidth}{!}{
\includegraphics[scale=1]{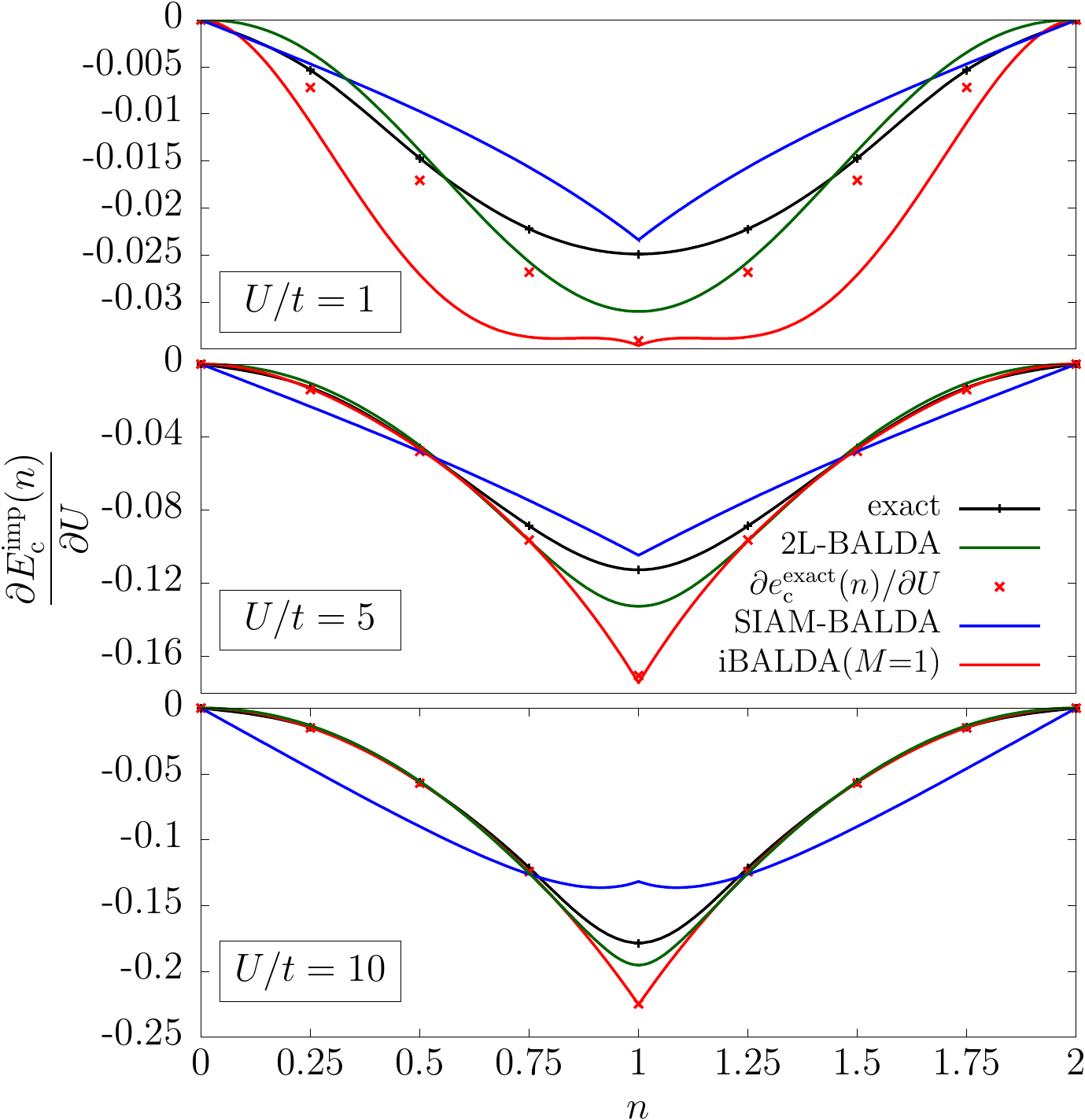}
}
\caption{
Single-impurity density-functional correlation energy
derivative with respect to $U$ calculated for various $U/t$ values. The
derivative shown for iBALDA($M$=1) is, by construction, the derivative
of the per-site BALDA correlation energy. Exact results were
obtained for the 8-site ring. See text and
Appendix~\ref{appendix:Lieb_max} for further details.  
}
\label{fig:dEcimp_dU}
\end{figure}

Returning to the single-impurity DFAs ($M=1$), it is quite instructive
to plot the derivative in $U$ of the various functionals
in order to analyze further the double occupations shown in Fig.~\ref{fig:dblocc}. According to
Eqs.~(\ref{eq:ecbath_per_site_M_uniform}) and (\ref{eq:dblocc_SOET}), 
both full $e_{\rm c}(n)$ and
impurity correlation energy derivatives should in
principle be analyzed. However, at half-filling, and for any $U$
and 
$t$ values, BALDA becomes exact for $e_{\rm c}(n)$ when
approaching the thermodynamic limit, by
construction~\cite{lima2003density}. As a result, approximations in the
impurity correlation functional will be the major source of errors which
are purely functional-driven in the half-filled case, since the exact
embedding potential is known (i.e. the correct uniform density profile
is obtained when solving Eq.~(\ref{eq:self-consistent-SOET_new}) in this case). Results are plotted with respect to the
density in
Fig.~\ref{fig:dEcimp_dU}, thus providing a clear picture not only at
half-filling but also around the latter density regime. The exact results obtained by Lieb
maximization 
for the 8-site ring are
used as reference [see the technical details in Appendix~\ref{appendix:Lieb_max}]. 
Let us recall that, within iBALDA($M$=1), the impurity correlation energy is
approximated with the full per-site
correlation one $e_{\rm c}(n)$, which is then modeled at the BALDA
level of approximation. The substantial (negative) difference between
the exact full per-site and impurity correlation energy derivatives at
half-filling (see Fig.~\ref{fig:dEcimp_dU}),
which is missing in iBALDA($M$=1), explains why the latter approximation
systematically overestimates the double occupation. Interestingly, the
derivative obtained with iBALDA($M$=1) at $n=1$, which is nothing but
the derivative of the BALDA per-site correlation energy at half-filling, is essentially
on top of its exact 8-site analog, thus confirming that finite-size
effects are negligible. Turning to SIAM-BALDA, the derivative in $U$ of
the impurity correlation energy turns out to be relatively accurate at $n=1$ for
$U/t=1$ and $U/t=5$, thus leading to good double occupations in this regime of
correlation. The derivative deteriorates for the larger $U/t=10$ value,
as expected~\cite{senjean2017site}. Note finally that 2L-BALDA, where
the impurity
correlation functional is approximated by its analog for the Hubbard
dimer, is the only approximation that provides reasonable derivatives in
all correlation and density regimes. The stronger the correlation is,
the more accurate the method is.\\

Let us finally discuss the per-site energies which have been computed according to
Eq.~(\ref{eq:per_site_ener_M-soet_final4}) for the various DFAs. Results
obtained at half-filling are shown in Fig.~\ref{fig:per-site_fctU_N32}. 
\begin{figure}
\resizebox{0.49\textwidth}{!}{
\includegraphics[scale=1]{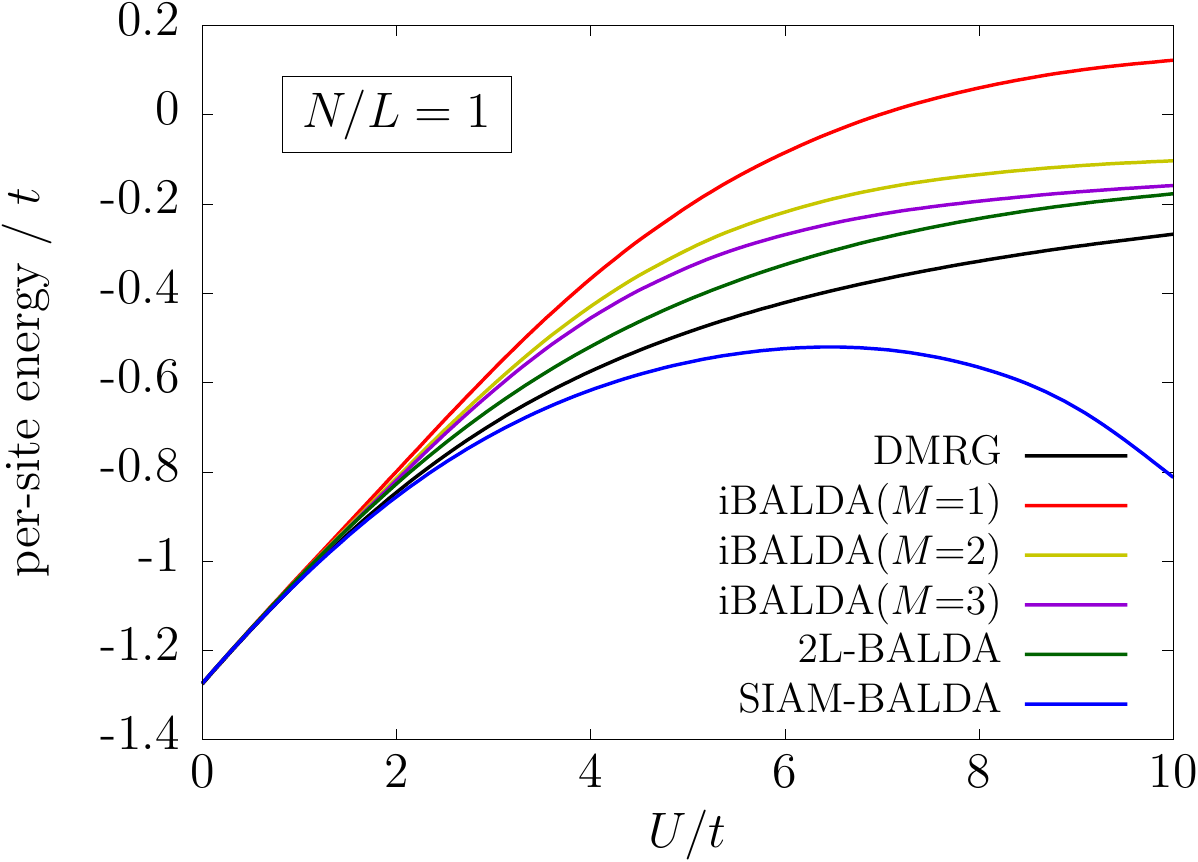}
}
\caption{
Per-site energies obtained from half-filled SOET calculations for various single-
and multiple-impurity DFAs. Standard DMRG 
results are used as reference. See text for further
details.
}
\label{fig:per-site_fctU_N32}
\end{figure}
The discussion on the performance of each DFA for the double occupation
turns out to hold also
for the energy. This is simply due to the fact that the per-site
non-interacting kinetic energy expression in Eq.~(\ref{eq:ts_fun}) is highly
accurate (it becomes exact in
the thermodynamic limit), like the BALDA per-site correlation energy at
half-filling (and therefore its derivative with respect to $t$). It then
becomes clear, when comparing Eqs.~(\ref{eq:dblocc_SOET}) and
(\ref{eq:per_site_ener_M-soet_final4}), that, at half-filling, the only source of errors in
the per-site energy is, like in the double occupation, the derivative in $U$ of the impurity correlation
functional. 

\subsection{Functional-driven errors away from half-filling}\label{subsec:fun_driv_error}

Away from half-filling, the exact embedding potential is not uniform 
anymore in the bath~\cite{senjean2017site}, thus reflecting 
the dependence in the bath site occupations of the impurity correlation
energy or, equivalently, of the per-site bath
correlation energy [see
Eqs.~(\ref{eq:self-consistent-SOET_new}),~(\ref{eq:ecbath_per_site_M}),
and (\ref{eq:Ecbath=sum_ec+Mecbath})]. Such a dependence is neglected in
all the DFAs used in this work, which induces errors in the density when
the impurity-interacting wavefunction is computed self-consistently
according to Eq.~(\ref{eq:self-consistent-SOET_new}). This generates 
so-called density-driven errors in the calculation of both the energy
and the
double occupation~\cite{kim2013understanding}. The latter are analyzed
in detail in Sec.~\ref{subsec:SC_results}.\\ 

In this section, we focus on
the functional-driven errors. In other words, all density-functional
contributions are calculated with the exact uniform density, like in
Sec.~\ref{subsec:half-filling}. The corresponding per-site energies are
shown in Fig.~\ref{fig:per-site_fctn_zoom}. Only the most challenging range
of fillings (i.e. $0.6\leqslant N/L\leqslant 1$~\cite{senjean2017site}) is shown. 
\begin{figure}
\resizebox{0.49\textwidth}{!}{
\includegraphics[scale=1]{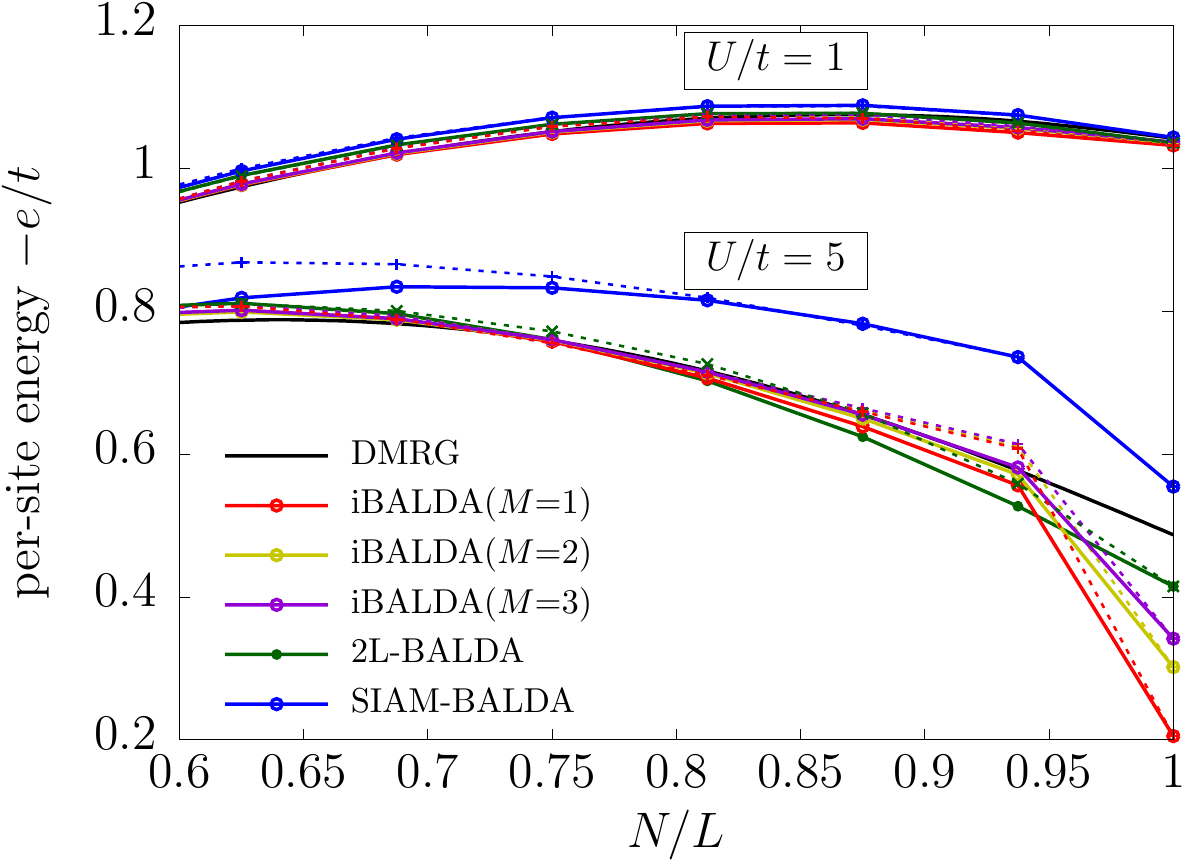}
}
\caption{Per-site energies calculated in SOET 
for fillings in the range $0.6\leqslant N/L\leqslant 1$ with
$U/t=1$ (upper curves) and $U/t=5$ (lower curves). Results obtained for
various DFAs with
exact (full lines) and self-consistently converged (dashed lines)
densities are shown. The iBALDA($M$=1) and reference DMRG results are
taken from Ref.~\cite{senjean2017site}. See text for further details.}
\label{fig:per-site_fctn_zoom}
\end{figure}
It clearly appears that iBALDA($M$=1), while failing dramatically in the
half-filled strongly correlated regime, performs relatively well away
from half-filling in all correlation regimes. It is the best
approximation in this regime of density. Increasing the number of
impurities has only a substantial effect on the energy when approaching
half-filling for large $U/t$ values. SIAM-BALDA performs reasonably well
in the weakly correlated regime but not as well as the other
functionals. In the same regime of correlation, 2L-BALDA stands in
terms of accuracy between iBALDA and SIAM-BALDA in the lower density
regime while giving the best result when approaching half-filling. The
latter statement holds also in the strongly correlated regime.\\

In order to further analyze the performance of each functional, let us
consider the derivative in $U$ of the single-impurity correlation functional
shown in Fig.~\ref{fig:dEcimp_dU}.~Away from half-filling and in the
strongly correlated regime, the full per-site and impurity correlation
energies give the same derivative so that neglecting the last term in
the right-hand side of Eq.~(\ref{eq:per_site_ener_M-soet_final4}), which
is done in iBALDA($M$), is well justified. Interestingly, this feature is
well reproduced by 2L-BALDA, where the impurity correlation energy
obtained from the Hubbard dimer is combined with the per-site BALDA
correlation energy [compare 2L-BALDA with iBALDA($M$=1) curves in the
bottom panel of Fig.~\ref{fig:dEcimp_dU}]. Obviously, SIAM-BALDA does
not exhibit the latter feature which explains why it fails in
this regime of correlation and density. Let us finally stress that, since BALDA provides an
accurate description of the full per-site correlation energy in the
strongly correlated regime, the second term in the right-hand side of
Eq.~(\ref{eq:per_site_ener_M-soet_final4}) is expected to be well
described in all the DFAs considered in this work (since they all use
BALDA for this contribution). As shown in Fig.~\ref{fig:dEcimp_dt}, this is actually the case [$\partial
e^{\rm exact}_{\rm c}(n)/\partial t$ should be compared with the
iBALDA($M$=1) derivative].  
\begin{figure}
\resizebox{0.49\textwidth}{!}{
\includegraphics[scale=1]{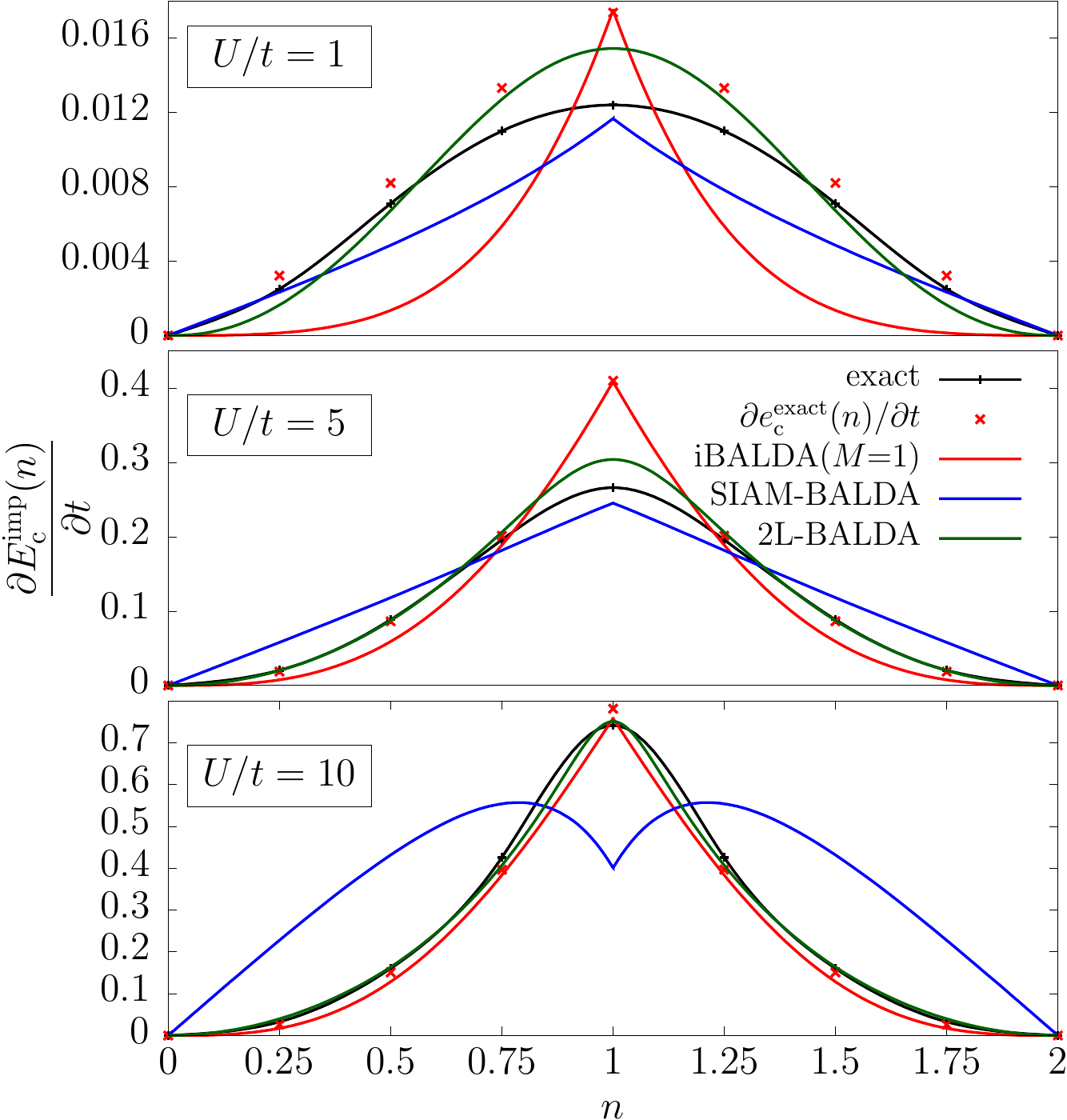}
}
\caption{
Single-impurity density-functional correlation energy
derivative with respect to $t$ calculated for various $U/t$ values. 
The
derivative shown for iBALDA($M$=1) is, by construction, the derivative
of the per-site BALDA correlation energy.
Exact results were
obtained for the 8-site ring. See text and
Appendix~\ref{appendix:Lieb_max} for further details.  
}
\label{fig:dEcimp_dt}
\end{figure}
\\

Turning to the weaker $U/t=1$ correlation regime, iBALDA($M$=1) underestimates the
derivative in $t$ of the full per-site correlation energy [compare $\partial
e^{\rm exact}_{\rm c}(n)/\partial t$ with 
iBALDA($M$=1) in Fig.~\ref{fig:dEcimp_dt}] away from half-filling while setting to zero the
derivative in $U$ of the per-site bath correlation functional whose
accurate value is actually negative [compare $\partial
e^{\rm exact}_{\rm c}(n)/\partial U$ with the exact impurity curve in
Fig.~\ref{fig:dEcimp_dU}]. The cancellation of errors leads
to the relatively accurate results shown in
Fig.~\ref{fig:per-site_fctn_zoom}. Turning to SIAM-BALDA in the same
density and correlation regime, the (negative) derivative
in $U$ of the per-site bath correlation energy is significantly
overestimated [compare iBALDA($M$=1) with SIAM-BALDA in the top panel of
Fig.~\ref{fig:dEcimp_dU}], thus giving a total per-site energy lower
than iBALDA($M$=1), as can be seen from the upper curves in 
Fig.~\ref{fig:per-site_fctn_zoom}. Interestingly, the latter derivative
in $U$ is less overestimated when using 2L-BALDA. This explains why it
performs better than SIAM-BALDA but still not as well as iBALDA($M$=1)
which benefits from error cancellations.

\subsection{Self-consistent results}\label{subsec:SC_results} 

We discuss in this section the results obtained by solving the density-functional
impurity problem self-consistently [see
Eq.~(\ref{eq:self-consistent-SOET_new})], thus accounting for not only
functional-driven but also density-driven
errors~\cite{kim2013understanding}. Self-consistently converged
densities obtained on the impurity site(s) for various DFAs are shown in
Figs.~\ref{fig:occ} and \ref{fig:occ_m3}.
\begin{figure}
\resizebox{0.49\textwidth}{!}{
\includegraphics[scale=1]{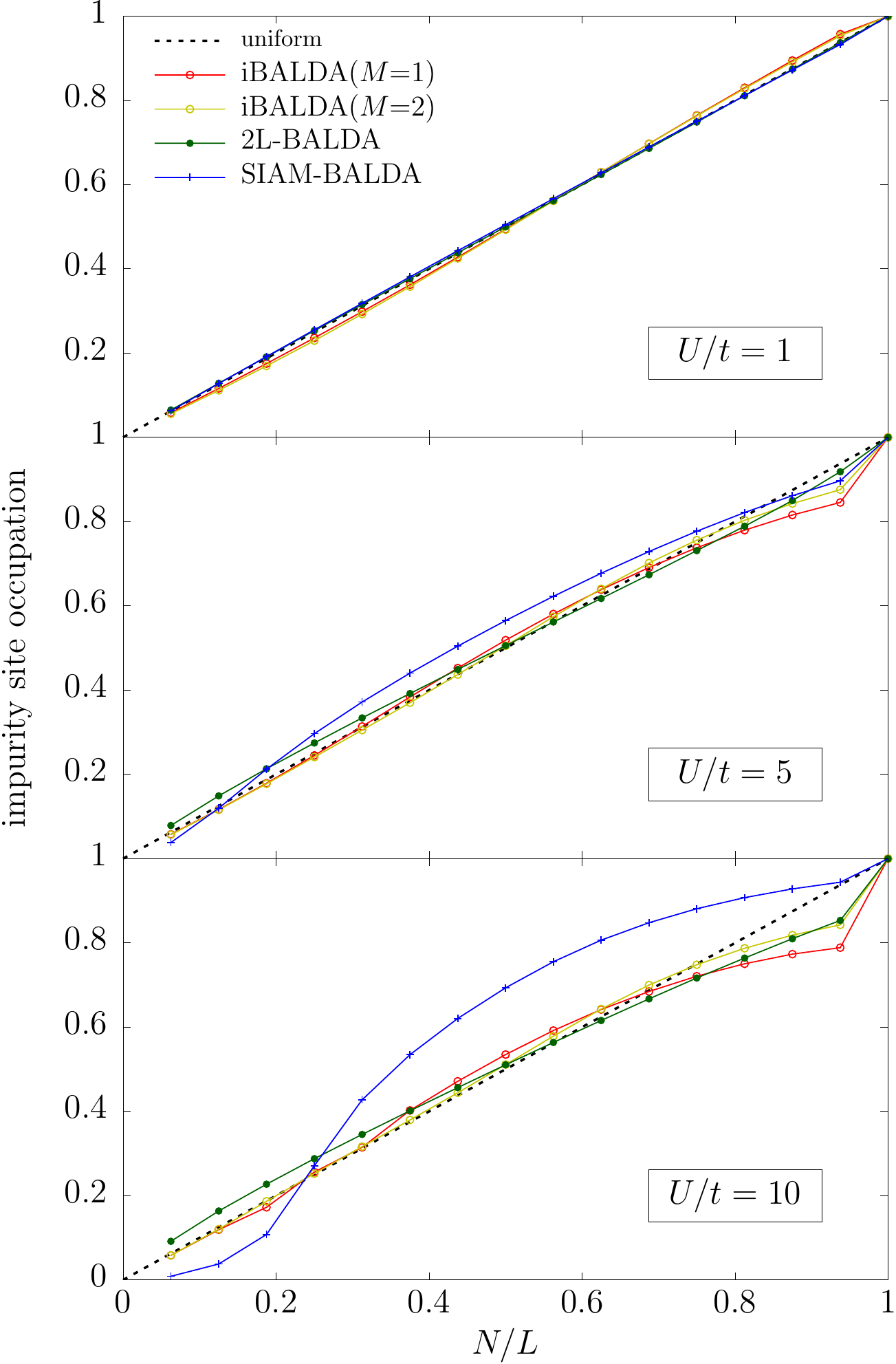}
}
\caption{
Self-consistently converged densities on the impurity site(s)
plotted against the filling for various DFAs and $U/t$ values. For
symmetry reasons, the two impurities in iBALDA($M$=2) have the same
density.}
\label{fig:occ}
\end{figure}
\begin{figure}
\resizebox{0.49\textwidth}{!}{
\includegraphics[scale=1]{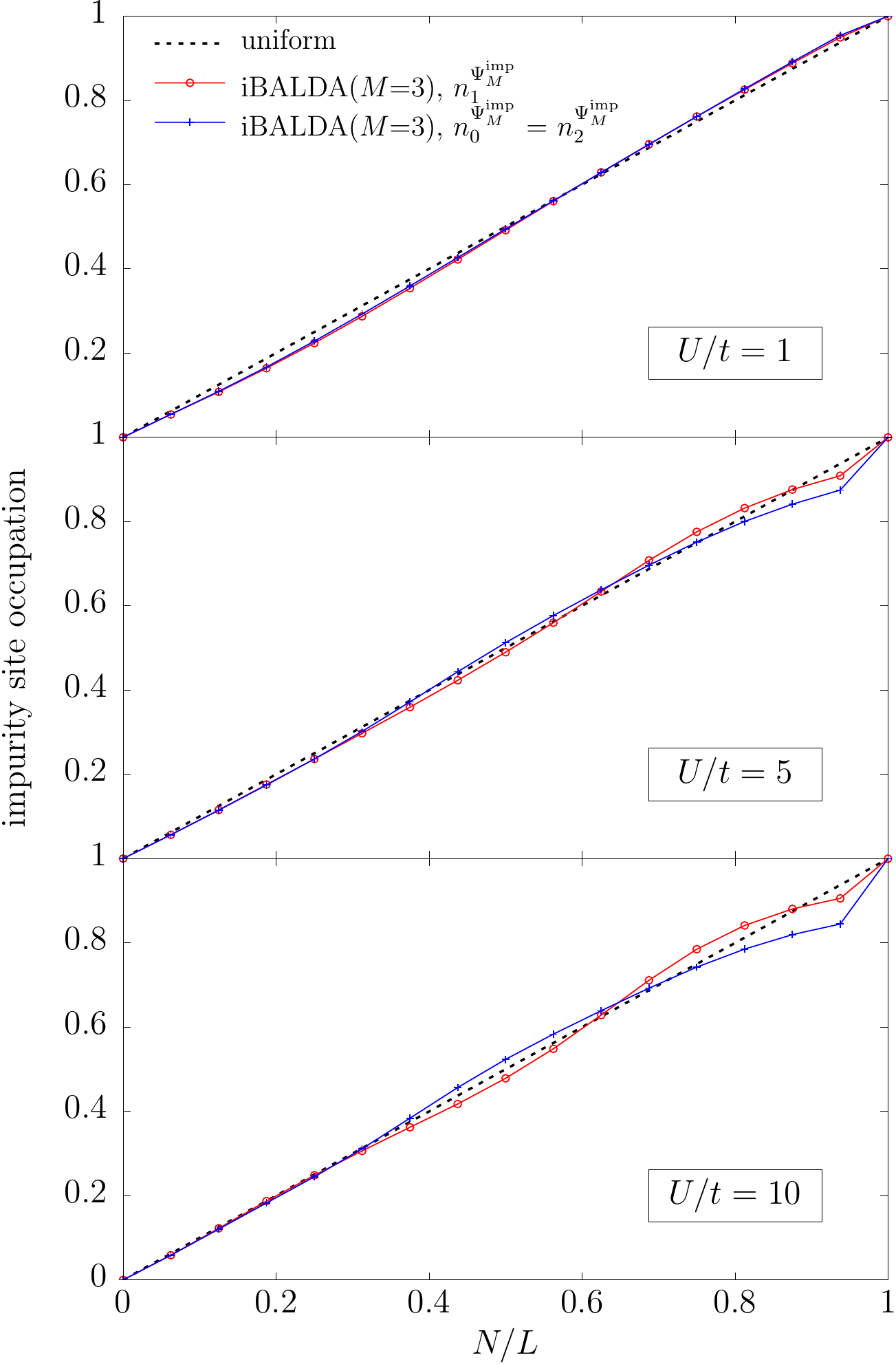}
}
\caption{
Self-consistently converged densities on the impurity sites
plotted against the filling for iBALDA($M$=3) with different $U/t$
values. The central and neighboring impurity densities are denoted as
$n_1^{\Psi^{\rm imp}_M}$ and 
$n_0^{\Psi^{\rm imp}_M}=n_2^{\Psi^{\rm imp}_M}$, respectively.}
\label{fig:occ_m3}
\end{figure}
Note that, at half-filling ($n=1$), the exact embedding potential has
been used, thus providing, for this particular density, the exact uniform
density profile. Away from half-filling, approximate density-functional
embedding potentials have been used, thus giving a density profile
that is not strictly uniform anymore. Interestingly, for all the DFAs except
SIAM-BALDA, the deviation from uniformity is more pronounced when
approaching the half-filled strongly correlated regime. In the case of
iBALDA, we notice that errors in the density are attenuated when
increasing the number of impurities, as expected. On the other hand, 
SIAM-BALDA generates huge density errors for almost all fillings when
the correlation is strong.\\

Turning to the weaker $U/t=1$ correlation
regime, we note that, in contrast to iBALDA, both self-consistent SIAM-BALDA and
2L-BALDA calculations hardly break the exact uniform density
profile (see the top panel of Fig.~\ref{fig:occ}). 
This can be rationalized by plotting the various impurity correlation
potentials (see Fig.~\ref{fig:dEcimp_dn}). 
\begin{figure}
\resizebox{0.49\textwidth}{!}{
\includegraphics[scale=1]{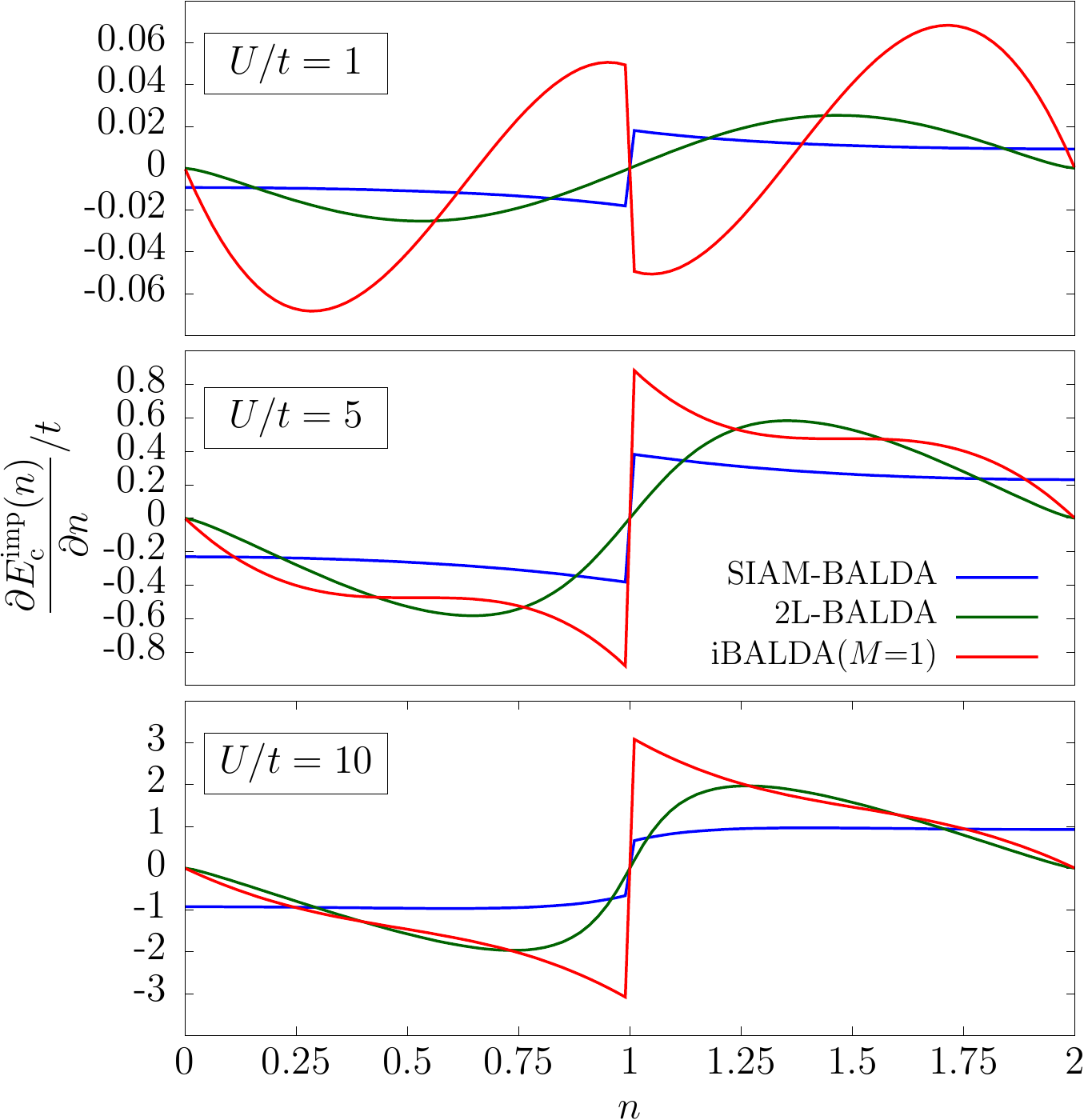}
}
\caption{Single-impurity density-functional correlation potential
obtained for various DFAs and $U/t$ values. The iBALDA($M$=1) potential
is, by construction, the BALDA correlation potential.}
\label{fig:dEcimp_dn}
\end{figure}
Let us first stress that, for all the DFAs
considered in this work, the correlation contribution to the embedding
potential on site $i$ reads [see
Eqs.~(\ref{eq:self-consistent-SOET_new}), (\ref{eq:Ecbath_expression})
(\ref{eq:lda_full_corr_ener}), and (\ref{eq:ilda_M-imp})]
\be\label{corr_emb_pot}
\dfrac{\partial e_{\rm
c}(n_i)}{\partial n_i}-\sum^{M-1}_{j=0}\delta_{ij}\dfrac{\partial E^{\rm imp}_{{\rm
c},M}(n_j)}{\partial n_j}.
\ee
The latter
potential will therefore be equal to zero on the impurity sites if iBALDA
is used. It will be uniform and equal to $\partial
e_{\rm c}(n)/\partial n$ (at least in the first iteration of the
self-consistent procedure as we start with a uniform density profile) in
the bath. 
As shown in the top
panel of Fig.~\ref{fig:dEcimp_dn} [see the iBALDA($M$=1) curve], in the
weakly correlated regime, the
latter potential, which is described with BALDA, is strongly
attractive for densities lower than 0.6. 
This
will induce a depletion of the density on the impurity sites, as clearly
shown in the top panels of Figs.~\ref{fig:occ} and \ref{fig:occ_m3}. The
opposite situation is observed for 
densities in the range $0.6\leqslant n\leqslant 1$, as expected from the strongly repulsive 
character of the potential. Note that this charge transfer process, which is completely
unphysical~\cite{akande2010electric}, is due to the incorrect linear
behavior in $U$ of the BALDA correlation potential away from the
strongly correlated regime~\cite{senjean2017site}. As readily seen from the first-order
expansion in $U$ given in Eq.~(32) of
Ref.~\cite{senjean2017site}, the latter potential is expected to change
sign at $n=(2/\pi)\arcsin\left(8/\pi^2\right)\approx 0.6$, which is in
complete agreement with the top panel of Fig.~\ref{fig:dEcimp_dn}.
Returning to Eq.~(\ref{corr_emb_pot}), for the single-impurity DFAs
SIAM-BALDA and 2L-BALDA, the correlation contribution to the embedding
potential reads $\partial
e_{\rm c}(n)/\partial n-\partial E_{{\rm c}}^{\rm imp}(n_0)/\partial
n_0$ on the impurity site and still $\partial
e_{\rm c}(n)/\partial n$ in the bath.
As shown in the top panel of
Fig.~\ref{fig:dEcimp_dn}, SIAM-BALDA and 2L-BALDA 
impurity correlation potentials dot not deviate significantly from zero
and, unlike iBALDA,
they do not exhibit unphysical features in the weakly correlated
regime, which explains why both DFAs give relatively good densities.
Turning to iBALDA($M$=3) in the strongly correlated regime (middle and
bottom panels of Fig.~\ref{fig:occ_m3}) and in the vicinity of half-filling,
the impurity site 1 better reproduces
the physical occupation $N/L$ than its nearest impurity neighbors (sites 0 and 2).
It can be explained as follows. Site 1 is the central site of the 
($M$=3)-impurity-interacting fragment, while sites 0 and 2 are 
directly connected to the (non-interacting and therefore unphysical) bath.
As a consequence, site 1 ``feels'' the bath less
than sites 0 and 2. It is, like in the physical system, surrounded by 
interacting sites. Interestingly, a similar
observation has been made by
Welborn \etal~\cite{welborn2016bootstrap}
within the {\it Bootstrap-Embedding method},
which is an improvement of DMET regarding the interaction between the fragment edge and the bath.
\\

Let us now refocus on the poor performance of SIAM-BALDA in the strongly
correlated regime. As shown in the middle and bottom panels of
Fig.~\ref{fig:dEcimp_dn}, the correlation part of the embedding
potential on the impurity site [which corresponds to the difference
between iBALDA($M$=1) and SIAM-BALDA curves] is, in this case, repulsive for densities
lower than about 0.25 and strongly attractive in the range $0.25\leqslant
n\leqslant 1$. The latter observation explains the large increase of density
on the impurity site in the self-consistent procedure, as depicted in
the bottom panel of Fig.~\ref{fig:occ}.\\ 

Finally, as shown in Fig.~\ref{fig:per-site_fctn_zoom}, using
self-consistently converged densities rather than exact (uniform)
ones can induce substantial density-driven errors on the
per-site energy, especially in the strongly correlated regime.
Interestingly, both functional- and density-driven errors somehow
compensate for 2L-BALDA around half-filling (see the lower curves in Fig.~\ref{fig:per-site_fctn_zoom}). This is not the case
anymore exactly at half-filling since we used the exact embedding potential,
thus removing density-driven errors completely. Regarding SIAM-BALDA,
the large error in the converged density obtained for $N/L=0.6$ at
$U/t=5$ (see the middle panel of Fig.~\ref{fig:occ}) is reflected on the
per-site energy. In this case, errors just accumulate. For larger
fillings, SIAM-BALDA gives a better density on the impurity site but the
functional-driven errors remain substantial.  

\subsection{Derivative discontinuity at half-filling}\label{subsec:DD_imp_corr_pot}    

As illustrated in Ref.~\cite{senjean2017site} in the special case of
the atomic limit (i.e. $t = 0$ or, equivalently, $U \rightarrow
+\infty$), the impurity correlation potential on the impurity site should
undergo a discontinuity at half-filling in the thermodynamic limit (even
for finite values of $U$). This
feature has fundamental and practical implications, in particular for
the calculation of the fundamental gap. The latter quantity plays a
crucial role in the description of the metal-insulator transition
which, in the one-dimensional Hubbard model, appears as soon as the on-site repulsion is
switched on ($U > 0$)~\cite{NoMott_Hubbardmodel}.\\

As shown in Fig.~\ref{fig:dEcimp_dn}, it is present in the iBALDA and
SIAM-BALDA functionals, as a consequence of the hole-particle symmetry
condition used in their construction. For the former functional, the feature is simply inherited
from BALDA. On the other hand, even though the 2L-BALDA impurity correlation
potential, which is extracted from the Hubbard dimer,
also fulfills the particle-hole symmetry condition,
it smoothly tends to 0 when approaching $n=1$.
As a consequence, the potential exhibits no derivative discontinuity (DD) at
half-filling when $U/t$ is {\it finite}. 
It only does when $U/t\rightarrow +\infty$~\cite{senjean2017local}.
As discussed by Dimitrov \etal~\cite{dimitrov2016exact}, this is due
to the fact that the dimer functional reproduces an intra-system steepening
and not an inter-system DD. In other words, the change in density
in the functional does not correspond to a 
change in the total number of electrons. The latter is indeed fixed to 2
in the dimer. Only the number of electrons on the impurity site varies. 
The problem becomes equivalent to describing an inter-system DD only
when the impurity can be treated as an isolated system, which is indeed
the case in the atomic (or, equivalently, strongly correlated) limit.\\

Note finally that, from a practical point of view, exhibiting a DD is not necessarily an
advantage as convergence problems may occur around half-filling~\cite{lima2003density}. 
In this work, this problem has been bypassed by using the exact
embedding potential at half-filling.
Also, given that only 32 sites are considered, the closest uniform
occupation to half-filling
is obtained for 30 electrons, i.e. $n = 0.9375$, which is
far enough from the strictly half-filled situation.
Convergence issues are expected to arise
when approaching the
thermodynamic limit since the density can then be much closer to 1. The
practical solutions to this problem, which have been
proposed for conventional KS calculations and use either a finite temperature~\cite{xianlong2012lattice} 
or {\it ad-hoc} 
parameters~\cite{kurth2010dynamical,karlsson2011time,
ying2014solving}, could in principle be implemented in SOET. This is left for future work.
Note finally that, despite
the absence of DD in the 2L-BALDA impurity correlation potential,
the BALDA correlation potential is still employed for the bath within
2L-BALDA, which means that convergence issues will appear as soon as
occupations in the bath are close to 1.

\section{Conclusions and perspectives}\label{sec:conclu}

Several extensions of a recent work~\cite{senjean2017site} on {\it site-occupation embedding theory}
(SOET) for model Hamiltonians have been explored.
Exact expressions for per-site energies and double occupations have been
derived for an arbitrary number of impurity sites. A simple $M$-impurity embedding
density-functional approximation (DFA) based on the {\it Bethe ansatz local density
approximation} (BALDA) and referred to as iBALDA($M$) has been proposed and tested on the
one-dimensional Hubbard Hamiltonian. A new single-impurity DFA [referred
to as 2L-BALDA] which
combines BALDA with the impurity correlation functional of the two-level
(2L) Hubbard
system~\cite{senjean2017local} has also been proposed.
Finally, the performance of an existing
DFA based on the {\it single impurity Anderson model} (SIAM) and BALDA~\cite{senjean2017site},
hence its name SIAM-BALDA, has been analyzed in further details in all
correlation regimes. Both functional- and density-driven errors have
been scrutinized.
Among all the single-impurity DFAs, 2L-BALDA is clearly the one that
performs the best in all density and correlation regimes.       
Unfortunately, the convergence of the (too) simple iBALDA approximation
in the number of impurities towards the accurate DMRG results was shown
to be slow. Better DFAs for multiple impurities are clearly needed. This
is left for future work. 
\\

SOET can be extended further in many directions. First of all,
substituting a Green function calculation for a many-body wavefunction
one like DMRG is expected to reduce the computational cost of the
method. From a formal point of view, this would also enable us to connect
SOET with the {\it dynamical mean-field theory} (DMFT) [see, for
example, Ref.~\cite{ayral2017dynamical}]. Note that the current
formulation of SOET is canonical. It would be interesting to remap
(density wise) the
original fully-interacting Hubbard problem onto an open impurity system, in the spirit
of the {\it density matrix embedding theory} (DMET). This may be
achieved, in principle exactly, by combining SOET with partition
DFT~\cite{elliott2010partition}, or, in a more approximate way,
by projecting the whole impurity-interacting problem in SOET
onto a smaller embedded subspace. A
Schmidt decomposition could be employed in the latter case~\cite{knizia2012density}.
Finally, an important step forward
would be the exploration of SOET in higher dimensions. Work is currently
in progress in all these directions.\\ 

Let us finally mention that the basic idea underlying SOET, which
consists in extracting site (or orbital) occupations from a
partially-interacting system, can be extended to an {\it ab initio} quantum
chemical Hamiltonian by using, for example, its simplified
seniority-zero
expression~\cite{richardson1963restricted,richardson1964exact,limacher2016new}.
This will be presented in a separate work.

\section*{Acknowledgments}

E.F. would like to dedicate this work to the memory of J\'{a}nos
G. \'{A}ngy\'{a}n. He would also like to thank Andreas Savin for
a fruitful discussion on the train from Middelfart to Copenhagen. B.S.
thanks D. Carrascal for taking the time 
to check the parameterization of his Hubbard dimer functional,
and M. Sauban\`{e}re, L. Mazouin, and K. Deur for fruitful discussions.
This work was funded by the Ecole Doctorale des Sciences Chimiques 222 (Strasbourg),
the
ANR (MCFUNEX project, Grant No. ANR-14-CE06-
0014-01), the ``Japon-Unistra'' network as well as the
Building of Consortia for the Development of Human
Resources in Science and Technology, MEXT, Japan for
travel funding.

\clearpage
\appendix

\section{Exact embedding potential at half-filling for multiple impurities}\label{sec:hole_particle_sym}

Let us consider any density $\mathbf{n}\equiv\lbrace n_i\rbrace_i$ summing up to a number
$N=\sum_in_i$ of
electrons. Under hole-particle symmetry, this density becomes
$(\underline{2} - \mathbf{n})\equiv\lbrace 2-n_i\rbrace_i$ and the
number of electrons equals $2L-N$ where $L$ is the number of sites. We will prove that these two
densities give the {\it same} correlation energy for the $M$-impurity interacting system. 
Since, for any local potential $\mathbf{v}$, the variational principle
in Eq.~(\ref{ener_min_n}) reads as follows for an impurity-interacting
system,
\be
\mathcal{E}^{{\rm imp}}_M(\mathbf{v})=
\underset{\mathbf{n}}{\rm min} \Big \lbrace F^{\rm imp}_M(\mathbf{n}) + 
(\mathbf{v}| \mathbf{n}) \Big \rbrace,
\ee
which gives, for any density $\mathbf{n}$,
\be
F^{\rm imp}_M(\mathbf{n})\geq \mathcal{E}^{{\rm
imp}}_M(\mathbf{v})-(\mathbf{v}| \mathbf{n}),
\ee
thus leading to the Legendre--Fenchel transform expression,
\be\label{eq:LFtrans_M_imp}
F^{\rm imp}_M(\mathbf{n}) =  \underset{\mathbf{v}}{\rm sup} 
\Big\{ &&\mathcal{E}^{{\rm imp}}_M(\mathbf{v})  
- 
(\mathbf{v}|\mathbf{n})\Big\}.
\ee
By applying a hole-particle symmetry transformation to
Eq.~(\ref{eq:LFtrans_M_imp}) [we will now indicate the number of
particles in the impurity-interacting energies for clarity], we obtain
\begin{eqnarray}\label{eq:Fimp-hole}
F^{\rm imp}_M(\underline{2} - \mathbf{n}) =  \underset{\mathbf{v}}{\rm sup} 
\Big\{ \mathcal{E}^{{\rm imp},2L-N}_M(\mathbf{v})  - 2\sum_i v_i
+ 
(\mathbf{v}|\mathbf{n})\Big\}, \nonumber \\
\end{eqnarray}
where $\mathcal{E}^{{\rm imp},2L-N}_M(\mathbf{v})$ is the
($2L-N$)-particle ground-state of the following
$M$-impurity interacting Hamiltonian:
\begin{eqnarray}\label{eq:M-imp_int_H}
\hat{H}^{\rm imp}_M(\mathbf{v}) & = & -t \sum_{i\sigma} \left( \hat{c}_{i
\sigma}^\dagger \hat{c}_{i+1\sigma} + \mathrm{H.c.}\right) + \sum_{i\sigma}v_i 
\hat{c}_{i\sigma}^\dagger \hat{c}_{i\sigma} \nonumber \\
& & + U  \sum_{i=0}^{M-1}\hat{c}_{i\uparrow}^\dagger \hat{c}_{i\uparrow} 
\hat{c}_{i\downarrow}^\dagger \hat{c}_{i\downarrow}.
\end{eqnarray}
Applying the
hole-particle transformation to the creation and annihilation operators,
\begin{eqnarray}
\hat{c}_{i\sigma}^\dagger&\rightarrow&
\hat{b}_{i\sigma}^\dagger=(-1)^i\hat{c}_{i\sigma}
,
\nonumber\\ 
\hat{c}_{i\sigma}&\rightarrow&\hat{b}_{i\sigma}=(-1)^i\hat{c}_{i\sigma}^\dagger
,
\end{eqnarray}
to the
$M$-impurity-interacting Hamiltonian in Eq.~(\ref{eq:M-imp_int_H})
leads to
\begin{eqnarray}
\hat{H}^{\rm imp}_M(\mathbf{v})
&=& -t \sum_{i\sigma} \left( \hat{b}_{i
\sigma}^\dagger \hat{b}_{i+1\sigma} + \mathrm{H.c.}\right)
+ \sum_{i\sigma} v_i \hat{b}_{i\sigma}
\hat{b}_{i\sigma}^\dagger \nonumber \\
&&+ U  \sum_{i=0}^{M-1}\hat{b}_{i\uparrow}
\hat{b}_{i\uparrow}^\dagger  \hat{b}_{i\downarrow}
\hat{b}_{i\downarrow}^\dagger  ,
\end{eqnarray}
or, equivalently,
\begin{eqnarray}
\hat{H}^{\rm imp}_M(\mathbf{v})
&=& -t \sum_{i\sigma} \left( \hat{b}_{i
\sigma}^\dagger \hat{b}_{i+1\sigma} + \mathrm{H.c.}\right)
+2\sum_i v_i - \sum_{i\sigma} v_i \hat{b}_{i\sigma}^\dagger\hat{b}_{i\sigma}
 \nonumber \\
&&+ UM - U \sum_{i=0}^{M-1}\sum_\sigma \hat{b}_{i\sigma}^\dagger
\hat{b}_{i\sigma}
+ U \sum_{i} \hat{b}_{i\uparrow}^\dagger \hat{b}_{i\uparrow}
\hat{b}_{i\downarrow}^\dagger  \hat{b}_{i\downarrow} .\nonumber \\
\end{eqnarray}
Then, by substituting and shifting the potential as follows, 
\begin{eqnarray}\label{eq:shift_pot_LF}
\tilde{v}_i =
- v_i - U \sum_{j=0}^{M-1} \delta_{ij}
\end{eqnarray}
we finally obtain 
\begin{eqnarray}\label{eq:M-imp_int_H_shift}
\hat{H}^{\rm imp}_M(\mathbf{\tilde{v}})
&=& -t \sum_{i\sigma} \left( \hat{b}_{i
\sigma}^\dagger \hat{b}_{i+1\sigma} + \mathrm{H.c.}\right)
+2\sum_i v_i 
+ \sum_{i\sigma} \tilde{v}_i \hat{b}_{i\sigma}^\dagger\hat{b}_{i\sigma}
 \nonumber \\
&&+ UM
+ U \sum_{i} \hat{b}_{i\uparrow}^\dagger \hat{b}_{i\uparrow}
\hat{b}_{i\downarrow}^\dagger  \hat{b}_{i\downarrow}.
\end{eqnarray}
As readily seen from Eqs.~(\ref{eq:M-imp_int_H}) and (\ref{eq:M-imp_int_H_shift}),
the $(2L-N)$-electron ground-state energy 
$\mathcal{E}^{{\rm imp},2L-N}_M(\mathbf{v})$
of $\hat{H}^{\rm imp}_M(\mathbf{v})$ is connected to
the $N$-electron ground-state energy 
$\mathcal{E}^{{\rm imp},N}_M(\mathbf{\tilde{v}})$ of
$\hat{H}^{\rm imp}_M(\mathbf{\tilde{v}})$ by
\begin{eqnarray}\label{eq:impE-N-2L-Nrelation}
 \mathcal{E}^{{\rm imp},2L-N}_M
(\mathbf{v}) = \mathcal{E}^{{\rm imp},N}_M(\mathbf{\tilde{v}}) + 2 \sum_i v_i + MU.
\end{eqnarray}
Introducing Eq.~(\ref{eq:impE-N-2L-Nrelation}) into
Eq.~(\ref{eq:Fimp-hole}) leads to
\begin{eqnarray}\label{eq:Fimp_2-n_n}
F^{\rm imp}_M(\underline{2} - \mathbf{n}) & = & 
\underset{\mathbf{v}}{\rm sup} 
\left \lbrace \mathcal{E}^{{\rm imp},N}_M(\tilde{\mathbf{v}})  
+(\mathbf{v}|\mathbf{n})\right  \rbrace +MU\nonumber  \\
&=&
\underset{\tilde{\mathbf{v}}}{\rm sup} 
\left \lbrace \mathcal{E}^{{\rm imp},N}_M(\tilde{\mathbf{v}})  
-(\tilde{\mathbf{v}}|\mathbf{n})\right  \rbrace 
+U\left(M-\sum_{i=0}^{M-1}n_i\right)\nonumber  \\
&=& F^{{\rm imp}}_M(\mathbf{n})+U\left(M-\sum_{i=0}^{M-1}n_i\right).
\end{eqnarray}
Note that
the maximising potential in
Eq.~(\ref{eq:Fimp_2-n_n}), denoted by
$\tilde{v}^{\rm emb}_M(\mathbf{n})$, is nothing but the exact
embedding potential $v^{\rm emb}_M(\mathbf{n})$ which restores the
exact density profile $\mathbf{n}$, by definition:
\begin{eqnarray}\label{eq:tildev=v}
\tilde{v}^{\rm emb}_{M,i}(\mathbf{n}) = v^{\rm emb}_{M,i}(\mathbf{n}).
\end{eqnarray}
According to the shift in Eq.~(\ref{eq:shift_pot_LF}), 
this maximising potential is related to the maximising one in Eq.~(\ref{eq:Fimp-hole}),
denoted by $\mathbf{v}^{\rm emb}_M(\underline{2} - \mathbf{n})$, by
\begin{eqnarray}
\tilde{v}_{M,i}^{\rm emb}(\mathbf{n}) = - v^{\rm emb}_{M,i}(\underline{2} - \mathbf{n}) - U \sum_{j=0}^{M-1} \delta_{ij}.
\end{eqnarray}
From the equality~(\ref{eq:tildev=v}), it comes
\begin{eqnarray}
{v}_{M,i}^{\rm
emb}(\underline{2}-\underline{n})
=
-{v}^{\rm emb}_{M,i}(\underline{n}) -U \sum_{j=0}^{M-1} \delta_{ij}
\end{eqnarray}    
thus leading to, at half-filling,
\begin{eqnarray}
{v}^{\rm emb}_{M,i}(\underline{1})=
- \dfrac{U}{2} \sum_{j=0}^{M-1} \delta_{ij}.
\end{eqnarray}

\section{Fundamental relation between derivatives in $t$ and $U$ of the complementary bath per-site
correlation energy for multiple impurities}\label{appendix:fun_rel}

If we denote $\mathbf{v}_M^{\rm emb}(\mathbf{n})$ the maximizing
potential in the Legendre--Fenchel transform of 
Eq.~(\ref{eq:LFtrans_M_imp}), we deduce from the linearity in $t$ and
$U$ of the impurity-interacting Hamiltonian 
that [the dependence in $t$ and $U$ is now
introduced for clarity]
\be
F^{\rm imp}_M(t,U,\mathbf{n})=\Bigg[&&t
\dfrac{\partial \mathcal{E}^{{\rm
imp}}_M(t,U,\mathbf{v})}{\partial t}
\nonumber\\
&&+U
\dfrac{\partial \mathcal{E}^{{\rm
imp}}_M(t,U,\mathbf{v})}{\partial U}
\left.\Bigg]
\right|_{\mathbf{v}=\mathbf{v}_M^{\rm emb}(\mathbf{n})},
\ee     
thus leading to the fundamental relation
\be
F^{\rm imp}_M(t,U,\mathbf{n})&=&
t\dfrac{\partial F^{\rm imp}_M(t,U,\mathbf{n})}{\partial t}
\nonumber\\
&&+
U\dfrac{\partial F^{\rm imp}_M(t,U,\mathbf{n})}{\partial U},
\ee
as a consequence of the stationarity condition fulfilled by $\mathbf{v}_M^{\rm
emb}(\mathbf{n})$. Since both the non-interacting kinetic energy [which
is obtained when $U=0$]
and the impurity Hx functional [first term in the right-hand side of
Eq.~(\ref{eq:Hx_plus_c_imp})] fulfill the same relation, we conclude
from the decomposition in Eq.~(\ref{eq:KS_decomp_Mimp}) that 
\be\label{eq:fun_rel_Ecimp-M}
E^{\rm imp}_{{\rm c},M}(t,U,\mathbf{n})&=&
t\dfrac{\partial E^{\rm imp}_{{\rm c},M}(t,U,\mathbf{n})}{\partial t}
\nonumber\\
&&+
U\dfrac{\partial E^{\rm imp}_{{\rm c},M}(t,U,\mathbf{n})}{\partial U}.
\ee
We finally obtain, by combining 
Eqs.~(\ref{eq:ecbath_per_site_M}),~(\ref{eq:fun_rel_Ecimp-M}) and
(\ref{eq:fun_rel_ec}), the  
fundamental relation in Eq.~(\ref{eq:fun_rel_ecbath-M-imp}). 

\section{Lieb maximization and correlation energy derivatives for a single impurity}\label{appendix:Lieb_max}

The impurity-interacting LL functional in Eq.~(\ref{eq:LL_Mimp}) [we
consider the particular case of a single impurity ($M=1$) in
the following] can be
rewritten as a Legendre--Fenchel transform
~\cite{fromager2015exact,senjean2017local},
\begin{eqnarray}\label{eq:LF_single_imp}
F^{\rm imp}(t,U,\mathbf{n}) = \underset{\mathbf{v}}{\rm sup} \left\lbrace
\mathcal{E}^{\rm imp}(t,U,\mathbf{v}) - (\mathbf{v} | \mathbf{n}) \right\rbrace,
\end{eqnarray}
where $\mathcal{E}^{\rm imp}(t,U,\mathbf{v})$ is the ground-state energy
of $\hat{T}+U\hat{n}_{0\uparrow}\hat{n}_{0\downarrow}+\sum_i v_i
\hat{n}_i$. Note that the dependence in both $t$ and $U$ of $F^{\rm
imp}(\mathbf{n})$ and $\mathcal{E}^{\rm imp}(\mathbf{v})$ has been
introduced for clarity. The so-called Lieb maximization~\cite{LFTransform-Lieb} procedure
described in Eq.~(\ref{eq:LF_single_imp}) has been used in this work in
order to
compute accurate values of $F^{\rm
imp}(t,U,\mathbf{n})$ and $T_{\rm s}(t,\mathbf{n})=F^{\rm
imp}(t,U=0,\mathbf{n})$ for a 8-site ring. The impurity-interacting energy $\mathcal{E}^{\rm
imp}(t,U,\mathbf{v})$ has been obtained by performing an exact
diagonalization calculation based
on the Lanczos
algorithm~\cite{lanczos}. The impurity correlation energy is then obtained
as follows,   
\begin{eqnarray}\label{eq:Ec_imp_LF}
E_{\rm c}^{\rm imp}(t,U,\mathbf{n}) = F^{\rm imp}(t,U,\mathbf{n}) - T_{\rm
s}(t,\mathbf{n}) - \dfrac{U}{4}n_0^2.
\end{eqnarray}
Since $\partial F^{\rm imp}(t,U,\mathbf{n})/\partial U=d^{\rm
imp}(t,U,\mathbf{n})$ is the impurity site double occupation obtained for
the maximizing potential in Eq.~(\ref{eq:LF_single_imp}) [see
Eq.~(\ref{eq:sum_2ble_occ_fun}) and Eq.~(A5)
in Ref.~\cite{senjean2017site}], it comes from
Eq.~(\ref{eq:Ec_imp_LF}),
\be\label{eq:dEcimp_over_dU_fun}
\dfrac{\partial E_{\rm c}^{\rm imp}(t,U,\mathbf{n})}{\partial U}=d^{\rm
imp}(t,U,\mathbf{n})-\dfrac{n^2_0}{4}.
\ee
Moreover, since 
\be
&&t\dfrac{\partial F^{\rm imp}(t,U,\mathbf{n})}{\partial t}=T^{\rm
imp}(t,U,\mathbf{n})
\nonumber\\
&&=F^{\rm imp}(t,U,\mathbf{n})-Ud^{\rm
imp}(t,U,\mathbf{n})
\ee 
is the impurity-interacting kinetic energy
obtained for the maximizing potential in Eq.~(\ref{eq:LF_single_imp})
[see Eq.~(\ref{eq:Timp_uniform}) and Eq.~(B6)
in Ref.~\cite{senjean2017site}], which gives in the non-interacting case
$t\,\partial T_{\rm
s}(t,\mathbf{n})/\partial t=T_{\rm
s}(t,\mathbf{n})$, we recover from Eq.~(\ref{eq:Ec_imp_LF}) the
expression in Eq.~(B8) of Ref.~\cite{senjean2017site},  
\be
\dfrac{\partial E_{\rm c}^{\rm imp}(t,U,\mathbf{n})}{\partial t}
=\dfrac{T^{\rm
imp}(t,U,\mathbf{n})-T_{\rm
s}(t,\mathbf{n})
}{t},
\ee
which can be further simplified as follows, 
\be\label{eq:dEcimp_over_dt_simplified}
\dfrac{\partial E_{\rm c}^{\rm imp}(t,U,\mathbf{n})}{\partial t}
&=&\dfrac{
E_{\rm c}^{\rm imp}(t,U,\mathbf{n})}{t}
\nonumber
\\
&&+\dfrac{U}{t}\left[\dfrac{n_0^2}{4}-d^{\rm
imp}(t,U,\mathbf{n})\right].
\ee
Interestingly, the derivatives in $t$ and $U$ are connected as follows,
according to Eq.~(\ref{eq:dEcimp_over_dU_fun}),
\be\label{eq:dEcimp_dt_rel_dEcimp_dU}
\dfrac{\partial E_{\rm c}^{\rm imp}(t,U,\mathbf{n})}{\partial t}
&=&\dfrac{
E_{\rm c}^{\rm imp}(t,U,\mathbf{n})}{t}
\nonumber
\\
&&-\dfrac{U}{t}
\dfrac{\partial E_{\rm c}^{\rm imp}(t,U,\mathbf{n})}{\partial U}.
\ee
Thus we recover Eq.~(\ref{eq:fun_rel_Ecimp-M}) in the particular
single-impurity case.\\

Similarly, in the fully-interacting case, the LL functional can be
rewritten as
follows, as a consequence of Eq.~(\ref{ener_min_n}), 
\be\label{eq:LF_fully_inte}
F(t,U,\mathbf{n}) = \underset{\mathbf{v}}{\rm sup} \left\lbrace
{E}(t,U,\mathbf{v}) - (\mathbf{v} | \mathbf{n}) \right\rbrace,
\ee
where the $t$- and $U$-dependence in both $F(\mathbf{n})$ and $E(\mathbf{v})$ is now made
explicit. From the correlation energy expression, 
\be
E_{\rm c}(t,U,\mathbf{n})=F(t,U,\mathbf{n})-T_{\rm
s}(t,\mathbf{n})-\dfrac{U}{4}\sum_in_i^2,
\ee
and the expressions for the LL functional derivatives in $t$ and $U$ [those and their
above-mentioned impurity-interacting analogs are deduced from the Hellmann-Feynman theorem], 
\be\label{eq:dF_over_dU_fun}
\dfrac{\partial F(t,U,\mathbf{n})}{\partial
U}=\sum_id_i(t,U,\mathbf{n}),
\ee
and
\be\label{eq:dF_over_dt_fun}
t\dfrac{\partial F(t,U,\mathbf{n})}{\partial t}&=&T
(t,U,\mathbf{n})
\nonumber\\
&=&F(t,U,\mathbf{n})-U
\sum_id_i(t,U,\mathbf{n}),
\ee
it comes
\be\label{eq:dEc_over_dU_fun}
\dfrac{\partial E_{\rm c}(t,U,\mathbf{n})}{\partial
U}=\sum_id_i(t,U,\mathbf{n})-\dfrac{1}{4}\sum_in_i^2,
\ee 
and
\be\label{eq:dEc_over_dt_fun}
\dfrac{\partial E_{\rm c}(t,U,\mathbf{n})}{\partial
t}=\dfrac{T
(t,U,\mathbf{n})-T_{\rm s}
(t,\mathbf{n})}{t}.
\ee
Note that $d_i(t,U,\mathbf{n})$ and $T
(t,U,\mathbf{n})$, which have been introduced in
Eqs.~(\ref{eq:dF_over_dU_fun}) and~(\ref{eq:dF_over_dt_fun}),
denote the site $i$ double occupation and the total
(fully-interacting) kinetic energy,
respectively. Both are calculated for the maximizing potential
in Eq.~(\ref{eq:LF_fully_inte}).  
For a uniform density profile $\underline{n}$, the per-site correlation
energy
reads
\be
e_{\rm c}(t,U,n)
&=&\dfrac{E_{\rm c}(t,U,\underline{n})}{L}
\nonumber
\\
&=&\dfrac{
1}{L}
\Big(F(t,U,\underline{n})-T_{\rm
s}(t,\underline{n})\Big)-\dfrac{U}{4}n^2.
\ee
Since, in this case, $d_i(t,U,\underline{n})=d(t,U,n)$ is
site-independent, we finally obtain from Eqs.~(\ref{eq:dF_over_dt_fun}),~(\ref{eq:dEc_over_dU_fun})
and (\ref{eq:dEc_over_dt_fun}), 
\be\label{eq:decdU_uni_fun}
\dfrac{\partial e_{\rm c}(t,U,{n})}{\partial
U}=d(t,U,{n})-\dfrac{n^2}{4},
\ee
and
\be
\dfrac{\partial e_{\rm c}(t,U,{n})}{\partial
t}&=&\dfrac{F(t,U,\underline{n})
-T_{\rm s}
(t,\underline{n})}{tL}
-\dfrac{U}{t}d(t,U,{n}).
\nonumber\\
\ee
By analogy with Eq.~(\ref{eq:dEcimp_over_dt_simplified}), the latter
expression can be simplified as follows,
\be
\dfrac{\partial e_{\rm c}(t,U,{n})}{\partial
t}&=&\dfrac{
e_{\rm c}(t,U,{n})}{t}
+\dfrac{U}{t}\left[\dfrac{n^2}{4}-d(t,U,{n})\right],
\nonumber\\
\ee
or, equivalently (see Eq.~(\ref{eq:decdU_uni_fun})),
\be\label{eq:fun_rel_ec}
\dfrac{\partial e_{\rm c}(t,U,{n})}{\partial
t}&=&\dfrac{
e_{\rm c}(t,U,{n})}{t}
-\dfrac{U}{t}
\dfrac{\partial e_{\rm c}(t,U,{n})}{\partial
U}
.
\nonumber\\
\ee

\section{derivatives of BALDA}\label{sec:derivatives_BALDA}
\subsection{derivative with respect to $U$ and $t$}

\begin{widetext}
As readily seen in Eq.~(\ref{eq:dblocc_SOET}), the derivative of the complementary
bath per-site correlation energy functional with respect to $U$ is necessary to compute double occupation in SOET.
According to Eq.~(\ref{eq:ecbath_per_site_M_uniform}), it implies the derivative
of the conventional per-site correlation energy, modelled with BALDA,
which reads
\begin{eqnarray}
\dfrac{\partial e_{\rm c}^{\rm BALDA}(n \leqslant1, U/t)}{\partial U} =  \dfrac{\partial \beta (U/t)}{\partial U} \left[ \dfrac{-2t}{\pi} \sin \left( \dfrac{\pi n}{\beta (U/t)} \right) + \dfrac{2tn}{\beta (U/t)} \cos \left( \dfrac{\pi n}{\beta (U/t)}\right)\right] - \dfrac{n^2}{4},
\end{eqnarray}
and then for $n > 1$:
\begin{eqnarray}
\dfrac{\partial e_{\rm c}^{\rm BALDA}(n > 1, U/t)}{\partial U} =  \dfrac{\partial \beta (U/t)}{\partial U}
\left[ \dfrac{-2t}{\pi} \sin \left( \dfrac{\pi(2-n)}{\beta (U/t)} \right) + \dfrac{2t(2-n)}{\beta (U/t)} \cos \left( \dfrac{\pi(2-n)}{\beta (U/t)}\right)\right] + (n-1) - \dfrac{n^2}{4} \nonumber \\
\end{eqnarray}
where $\partial \beta(U/t) / \partial U = ( \partial \beta(U/t) / \partial (U/t) )/t$,
is computed with finite differences by solving Eq.~(\ref{eq:beta}) for $\beta(U/t)$.

The derivative with respect to $t$ are calculated according to Eq.~(\ref{eq:fun_rel_ec}).

\subsection{derivative with respect to $n$}

To get the correlation embedding potential, the derivatives of the 
correlation functionals with respect to $n$ is necessary.
The derivative of the convention per-site
density-functional correlation energy reads
\begin{eqnarray}
 \dfrac{\partial e_{\rm c}^{\rm BA}(n \leqslant1)}{\partial n} =  - 
 2t
 \cos \left( \dfrac{\pi n}{\beta(U/t)}\right) + 
 2t \cos \left( \dfrac{\pi n}{2} \right) - \dfrac{Un}{2},
\end{eqnarray}
and
\begin{eqnarray}
 \dfrac{\partial e_{\rm c}^{\rm BA}(n > 1)}{\partial n} = 
 2t
 \cos \left( \dfrac{\pi (2-n)}{\beta(U/t)}\right) -
 2t \cos \left( \dfrac{\pi (2-n)}{2} \right) + U - \dfrac{Un}{2}.
\end{eqnarray}

\section{derivatives of SIAM-BALDA}\label{sec:derivatives_SIAM-BALDA}

The derivatives of the SIAM-BALDA impurity correlation functional [Eq.~(\ref{eq:anderson_secondorder})] are given
with
respect to $U$ for $n \leqslant 1$ as follows,
\begin{eqnarray}
\dfrac{\partial E_{{\rm c}, U/\Gamma \rightarrow 0}^{\rm SIAM}(U,\Gamma(t,n)) }{\partial U} = - \dfrac{2 \times 0.0369}{\pi}\left( \dfrac{U}{\Gamma(t,n)}\right) + \dfrac{4\times 0.0008}{\pi^3}\left( \dfrac{U}{\Gamma(t,n)}\right)^3.
\end{eqnarray}
The derivative with respect to $t$ is given according to Eq.~(\ref{eq:dEcimp_dt_rel_dEcimp_dU}).
Then, the impurity correlation potential is determined
by the derivative of the functional with respect to the occupation number $n$:
\begin{eqnarray}
 \dfrac{\partial E_{{\rm c}, U/\Gamma \rightarrow 0}^{\rm SIAM}(U,\Gamma(t,n))}{\partial n} = \dfrac{\partial \Gamma(t,n)}{\partial n} \left.\dfrac{\partial E_{{\rm c}, U/\Gamma \rightarrow 0}^{\rm SIAM}(U,\Gamma)}{\partial \Gamma}\right|_{\Gamma = \Gamma(t,n)},
\end{eqnarray}
where
\begin{eqnarray}
\dfrac{\partial E_{{\rm c}, U/\Gamma \rightarrow 0}^{\rm SIAM}(U,\Gamma) }{\partial \Gamma} = \dfrac{0.0369}{\pi}
\left( \dfrac{U}{\Gamma}\right)^2
- \dfrac{3\times 0.0008}{\pi^3}\left( \dfrac{U}{\Gamma}\right)^4,
\end{eqnarray}
and
\begin{eqnarray}
\dfrac{\partial \Gamma(t,n)}{\partial n}  = t \left( \dfrac{-\dfrac{\pi}{2}\sin ^2 (\pi n /2) - (1 + \cos (\pi n /2)) \dfrac{\pi}{2}\cos(\pi n /2)}{\sin^2(\pi n /2)} \right) = - \dfrac{\pi t}{2} \left( \dfrac{1 + \cos (\pi n /2)}{\sin^2(\pi n /2)}  \right) = - \dfrac{\pi \Gamma (t,n)}{2\sin(\pi n /2)}.
\end{eqnarray}
If $n > 1$, the particle-hole formalism imposes to use 
$\Gamma(t,2-n)$ instead of $\Gamma(t,n)$. The derivatives with respect to $n$ should be
changed accordingly.

\section{derivatives of 2L-BALDA}\label{sec:derivatives_2L-BALDA}

\subsection{Parametrization of the correlation energy of the dimer}

In this section, we summarize the parametrization 
of the Hubbard dimer
correlation energy by Carrascal and 
co-workers~\cite{carrascal2015hubbard,carrascal2016corrigendum}, 
necessary
to understand the following derivations.
The equations coming from their paper are referred to as ($\& N)$, where $N$ is 
the number of the equation.
We start from the definition of the correlation energy, where $n$ is the
occupation of the site 0 and $u = U/2t$ is a dimensionless parameter,
\begin{eqnarray}\label{eq:Ec_burke}
E_{\rm c}^{\rm 2L}(U,n) = f(g,\rho)\Big|_{\begin{subarray}{l}
  g=g(\rho,u) \\
  \rho = | n-1 |
  \end{subarray}}
  - T_{\rm s}(n) - E_{\rm Hx}(U,n),
\end{eqnarray}
where 2L refers to ``two-level'', and
\begin{eqnarray}
T_{\rm s}(n) = -2t \sqrt{n(2-n)}, ~~~~ E_{\rm Hx}(U,n) = U\left(1 - n\left(1 - \dfrac{n}{2} \right)\right).
\end{eqnarray}
To account for particle-hole 
symmetry of the functional, the variable $\rho = | n-1 |$ is used
rather than $n$ directly.
We now simply follow the guidelines from Eq.($\& 102$) to ($\& 107$), leading to
\begin{eqnarray}
f(g, \rho) = -2t g + U h(g,\rho),
\end{eqnarray}
and
\begin{eqnarray}
h(g,\rho) = \dfrac{g^2\left(1 - \sqrt{1 - \rho^2 - g^2}\right) + 2\rho^2}{2(g^2 + \rho^2)}.
\end{eqnarray}
Then, they proposed a first approximation to $g(\rho,u)$, denoted by the label 0:
\begin{eqnarray}
g_0(\rho,u) = \sqrt{\dfrac{(1 - \rho)(1 + \rho(1 + (1+\rho)^3 u a_1(\rho,u)))}{1 + (1+\rho)^3 u a_2(\rho,u)}},
\end{eqnarray}
where
\begin{eqnarray}
a_i(\rho,u) = a_{i1}(\rho) + u a_{i2}(\rho), ~~~ i = 1, 2
\end{eqnarray}
and
\begin{eqnarray}
a_{21}(\rho) = \dfrac{1}{2} \sqrt{\dfrac{\rho(1-\rho)}{2}} , ~~~ 
a_{12}(\rho) = \dfrac{1}{2}(1-\rho), ~~~
a_{11}(\rho) = a_{21}\left(1 + \dfrac{1}{\rho} \right), ~~~
a_{22}(\rho) = \dfrac{a_{12}(\rho)}{2}.
\end{eqnarray}
Plugging $g=g_0(\rho,u)$ into $f(g,\rho)$ leads to the first 
parametrization of $E_{\rm c}^{2L}(n)$ in Eq.~(\ref{eq:Ec_burke}).
In this work, we implemented the more accurate parametrization, given
in Eq.($\& 114$)~\cite{carrascal2016corrigendum}:
\begin{eqnarray}\label{eq:g_1}
g_1(\rho,u) = g_0(\rho,u) + \left( u\left.\dfrac{\partial h(g,\rho)}{\partial g}\right|_{g = g_0(\rho,u)}
-1 \right) q(\rho, u),
\end{eqnarray}
and where $q(\rho,u)$ is given in Eq.~($\& 115$) 
by~\cite{carrascal2016corrigendum}:
\begin{eqnarray}
q(\rho, u) = \dfrac{(1-\rho)(1+\rho)^3 u^2 [ (3\rho/2 - 1 + \rho(1+\rho)^3
u a_2(\rho,u))a_{12}(\rho) - \rho(1 + (1 + \rho)^3 u a_1(\rho,u))a_{22}(\rho)]}{2 g_0(\rho, u) (1 + (1 + \rho)^3 u a_2(\rho,u) )^2}.
\end{eqnarray}
The accurate pametrization of $E_{\rm c}^{2L}(n)$ is obtained by plugging
this $g_1(\rho,u)$ into $f(g,\rho)$, instead of $g_0(\rho, u)$. \\
In order to obtain the impurity correlation energy, 
a simple scaling of the
interaction parameter $U$ has to be applied
on the conventional correlation energy, as demonstrated in Ref.~\cite{senjean2017local} and given in Eq.~(\ref{eq:dimer_corr}),
leading to
\begin{eqnarray}\label{eq:Ec_imp_par}
E_{\rm c}^{\rm imp, 2L}(U, n) = E_{\rm c}^{2L}(U/2,n) = f(g,\rho)\Big|_{\begin{subarray}{l}
  g=g(\rho,u/2) \\
  \rho = | n-1 |
  \end{subarray}}
  - T_{\rm s}(n) - E_{\rm Hx}(U/2,n).
\end{eqnarray}

\subsection{derivative with respect to $U$ and $t$}

We compute the derivative with respect to the dimensionless parameter 
$u = U/2t$. 
The $\rho$- and $u$- dependence of $g(\rho, u)$ will be omitted for 
readability. Besides, many functions will be introduced, aiming to make
the implementation and its numerical verification easier.
Starting with
\begin{eqnarray}\label{eq:dEcimp/dU}
\dU{E_{\rm c}^{\rm 2L}(n)} &=& \dfrac{1}{2t} \left.\dfrac{\partial f(g,\rho)}{\partial u}\right|_{\begin{subarray}{l}
  g=g(\rho,u) \\
  \rho = | n-1 |
  \end{subarray}} - \left(1 + n\left(\dfrac{1}{2}n - 1\right)\right),
\end{eqnarray}
the impurity correlation functional reads, according to Eq.~(\ref{eq:Ec_imp_par}),
\begin{eqnarray}\label{eq:dEcimp/dU}
\dU{E_{\rm c}^{\rm imp,2L}(n)} &=& \dfrac{1}{4t} \left.\dfrac{\partial f(g,\rho)}{\partial u}\right|_{\begin{subarray}{l}
  g=g(\rho,u/2) \\
  \rho = | n-1 |
  \end{subarray}} - \dfrac{1}{2}\left(1 + n\left(\dfrac{1}{2}n - 1\right)\right),
\end{eqnarray}
with
\begin{eqnarray}
\du{f(g,\rho)} = -2t\left( \du{g} - h(g,\rho) - u \times \du{h(g,\rho)}\right).
\end{eqnarray}
The derivative of $h(g,\rho)$ is quite easy, as its only $u$-dependence is contained in $g$, so that:
\begin{eqnarray}
\du{h(g,\rho)} = \du{g} \dfrac{\partial h(g, \rho)}{\partial g},
\end{eqnarray}
with
\begin{eqnarray}\label{eq:dh_dg}
\dfrac{\partial h(g,\rho)}{\partial g} = g\dfrac{g^4 + 3g^2\rho^2 
+ 2\rho^2\left(\rho^2 - 1 - Y(g,\rho)\right)}{2\left(g^2 + \rho^2\right)^2Y(g,\rho)},
\end{eqnarray}
where the function $Y(g,\rho) = \sqrt{1 - g^2 - \rho^2}$ has been introduced.
For the first approximation, $g = g_0$ and
\begin{eqnarray}\label{eq:dg0_du}
\du{g_0(\rho, u)} = \du{\sqrt{G(\rho, u)}} = \dfrac{\partial G(\rho, u)/\partial u}{2\sqrt{G(\rho, u)}},
\end{eqnarray}
where $G(\rho,u) = N(\rho,u)/D(\rho,u)$ and
\begin{eqnarray}
N(\rho,u)&=&(1 - \rho) \left[1 + \rho \left(1 + (1 + \rho)^3 u a_1(\rho, u)\right)\right],\\
D(\rho , u) &=& 1 + (1 + \rho)^3 u a_2(\rho , u).
\end{eqnarray}
Their respective derivative with respect to $u$ reads
\begin{eqnarray}
\du{N(\rho, u)} &=&  (1 - \rho) \rho (1 + \rho)^3 \left( a_1(\rho ,u) + u \du{ a_1(\rho, u)} \right), \\
\du{D(\rho,u)} &=& \left( 1 + \rho \right) ^3  \left( a_2(\rho ,u) + u \du{a_2(\rho, u)} \right),
\end{eqnarray}
with
\begin{eqnarray}
\du{a_{2}(\rho, u)} = a_{22}(\rho), ~~~~ 
\du{a_{1}(\rho, u)} = a_{12}(\rho).
\end{eqnarray}
Turning to the second approximation $g = g_1$ implemented in this work, one get from the derivative of Eq.~(\ref{eq:g_1}),
\begin{eqnarray}\label{eq:dg1_du}
\dfrac{\partial g_1}{\partial u} &= &\dfrac{\partial g_0}{\partial u} 
+ \left( \left. \dfrac{\partial h(g,u)}{\partial g} \right|_{g=g_0}
+ u\dfrac{\partial}{\partial u}\left(\left.\dfrac{\partial h(g,u)}{\partial g}\right|_{g=g_0}\right) \right) q(\rho,u)+ \left(u\left.\dfrac{\partial h(g,u)}{\partial g}\right|_{g=g_0} - 1 \right) 
\dfrac{\partial q(\rho,u)}{\partial u}.
\end{eqnarray}
For convenience, we introduce two functions $w(g, u)$ and $v(g,u)$ so that
\begin{eqnarray}
\dfrac{\partial}{\partial u} \left( \dfrac{\partial h(g,u)}{\partial g} \right) = \left(\du{w(g,u)}v(g,u) - w(g,u)\du{v(g,u)}\right)\Big/v(g,u)^2,
\end{eqnarray}
with
\begin{eqnarray}
w(g,u) &=& g\left[g^4 + 3g^2\rho^2 + 2\rho^2\left(\rho^2 - 1 - Y(g,\rho)\right)\right],
\\
v(g,u) &=& 2Y(g,\rho) \left(g^2 + \rho^2\right)^2,
\end{eqnarray}
and
\begin{eqnarray}
\du{w(g,u)} &= &\dfrac{\partial g}{\partial u} \left[ g^4 + 3g^2\rho^2 + 2\rho^2 \left(\rho^2 - 1 - Y(g,\rho)\right)
+ g\left(4g^3 + 6g\rho^2+  \dfrac{2\rho^2g}{Y(g,\rho)}  \right) \right], \\
\du{v(g,u)} &=&  g\left(g^2 + \rho^2\right) \dfrac{\partial g}{\partial u} \left[\dfrac{-2\left(g^2 + \rho^2\right)}{Y(g,\rho)} + 8Y(g,\rho)  \right].
\end{eqnarray}
Finally, the last term in Eq.~(\ref{eq:dg1_du}) reads, for $q(\rho,u) = j(\rho,u)k(\rho,u)/l(\rho,u)$:
\begin{eqnarray}
\dfrac{\partial q(\rho,u)}{\partial u} = \dfrac{\left(\du{j(\rho,u)}k(\rho,u) + j(\rho,u)\du{k(\rho,u)}\right)l(\rho,u) - j(\rho,u)k(\rho,u)\du{l(\rho,u)}}{l(\rho,u)^2}
\end{eqnarray}
with
\begin{eqnarray}
j(\rho,u) &=& (1-\rho)(1+\rho)^3 u^2, \\
k(\rho, u) &=& \left(3\rho /2 - 1 + \rho(1 + \rho)^3 u a_2 (\rho, u)\right) a_{12}(\rho) - \rho\left(1 + (1+\rho)^3 \lambda u a_1 (\rho, u)\right) a_{22}(\rho) \\
l(u) &=& 2g_0(u) \left[1 + (1+\rho)^3 \lambda u a_2 (\rho, u,\lambda) \right]^2,
\end{eqnarray}
and their derivative with respect the $u$:
\begin{eqnarray}
\du{j(\rho,u)} &=& 2(1-\rho)(1+\rho)^3\lambda u \\
\du{k(u)} &=& a_{12}(\rho)\left[ \rho(1+\rho)^3 \left( a_2 (\rho,u) 
+ u \dfrac{\partial a_2 (\rho,u)}{\partial u} \right) \right] 
- a_{22}(\rho) \left[\rho (1+\rho)^3 \left( a_1 (\rho,u) + u\dfrac{\partial a_1(\rho,u)}{\partial u}\right) \right] \\
\du{l(u)} &=& 4g_0(u) \left[1 + (1+\rho)^3  u a_2 (\rho,u)\right] (1+\rho)^3  \left(a_2(\rho,u) + u \dfrac{\partial a_2(\rho,u)}{\partial u} \right)  + 2\dfrac{\partial g_0}{\partial u} \left[1 + (1+\rho)^3 \lambda u a_2 (\rho,u) \right]^2.
\end{eqnarray}
The derivative with respect to $t$ is given according to Eq.~(\ref{eq:dEcimp_dt_rel_dEcimp_dU}).

\subsection{derivative with respect to $n$}

Regarding the derivative with respect to $n$ which is necessary to get the embedded
correlation potential, it comes
\begin{eqnarray}
\dfrac{\partial E_{\rm c}^{\rm imp,2L}(n)}{\partial n} =  \dfrac{\partial \rho}{\partial n}
\left.\dfrac{f(g,\rho)}{\partial \rho}\right|_{\begin{subarray}{l}
  g=g(\rho,u/2) \\
  \rho = | n-1 |
  \end{subarray}}
  - \dfrac{\partial \mathcal{T}_s(n)}{\partial n} + \dfrac{U}{2}(1-n)
\end{eqnarray}
where $\partial \rho / \partial n = {\text{sign($n-1$)}}$ and
\begin{eqnarray}
\dfrac{\partial \mathcal{T}_s(n)}{\partial n} = - \dfrac{2t(1 - n)}{\sqrt{n(2-n)}}.
\end{eqnarray}
We start with
\begin{eqnarray}\label{eq:df_drho}
\dfrac{\partial f(g,\rho)}{\partial \rho} = -2t\dfrac{\partial g}{\partial \rho} + U \dfrac{\partial h(g,\rho)}{\partial \rho},
\end{eqnarray}
where, for the first parametrization using $g = g_0(\rho,u)$,
\begin{eqnarray}
\dfrac{\partial g_0}{\partial \rho} & = & =\dfrac{1}{2g_0D(\rho,u)} \left( \dfrac{\partial N(\rho,u)}{\partial \rho} - g_0^2 \dfrac{\partial D(\rho,u)}{\partial \rho} \right),
\end{eqnarray}
with
\begin{eqnarray}
\dfrac{\partial N(\rho , u)}{\partial \rho} &=& -1 + (1 - 2\rho)\left(1+(1+\rho)^3ua_1(\rho , u)\right) + \rho u(1-\rho)(1+\rho)^2\left(3a_1(\rho , u) + (1+\rho)\dfrac{\partial a_1(\rho , u)}{\partial \rho}\right),\\
\dfrac{\partial D(\rho , u)}{\partial \rho} &=& u (1+\rho)^2\left(3a_2(\rho , u) + (1+\rho)\dfrac{\partial a_2(\rho , u)}{\partial \rho}\right),
\end{eqnarray}
and
\begin{eqnarray}
\dfrac{\partial a_1(\rho , u)}{\partial \rho} & = & \dfrac{\partial a_{11}(\rho)}{\partial \rho} + u\dfrac{\partial a_{12}(\rho)}{\partial \rho}, ~~~~~ 
\dfrac{\partial a_2(\rho , u)}{\partial \rho} = \dfrac{\partial a_{21}(\rho)}{\partial \rho} + u\dfrac{\partial a_{22}(\rho)}{\partial \rho} \\
\dfrac{\partial a_{12}(\rho)}{\partial \rho} &=&  2
\dfrac{\partial a_{22}(\rho)}{\partial \rho} = - \dfrac{1}{2}, ~~~~
\dfrac{\partial a_{21}(\rho)}{\partial \rho} = \dfrac{1 - 2\rho}{8\sqrt{(1-\rho)\rho/2}} , ~~~~ 
\dfrac{\partial a_{11}(\rho)}{\partial \rho} = \dfrac{\partial a_{21}(\rho)}{\partial \rho} \left( 1 + \dfrac{1}{\rho} \right) - \dfrac{1}{\rho^2} a_{21}(\rho).
\end{eqnarray}
Then, the right term in the right hand side of Eq.~(\ref{eq:df_drho}) is derived as:
\begin{eqnarray}
\dfrac{\partial h(g,\rho)}{\partial \rho} & = & \dfrac{1}{2(g^2 + \rho^2)} \left( 4\rho + 2g\dfrac{\partial g}{\partial \rho} \left( 1 - Y(g,\rho) \right) + g^2 \dfrac{g (\partial g/\partial \rho) + \rho}{Y(g,\rho)} \right)  - \dfrac{g(\partial g / \partial \rho) + \rho}{(g^2 + \rho^2)^2} \left( 2\rho^2 + g^2 \left(1 - Y(g,\rho) \right) \right). \nonumber \\
\end{eqnarray}
Turning to the second parametrization $g = g_1$, the derivative with respect to $\rho$ leads to
\begin{eqnarray}
\dfrac{\partial g_1}{\partial \rho} = \dfrac{\partial g_0}{\partial \rho} + \left(u\left.\dfrac{\partial h(g,\rho)}{\partial g}\right|_{g=g_0} - 1\right)\dfrac{\partial q(\rho,u)}{\partial \rho} + u\dfrac{\partial}{\partial \rho} \left(\left.\dfrac{\partial h(g,\rho)}{\partial g}\right|_{g=g_0}\right) q(\rho,u)
\end{eqnarray}
with
\begin{eqnarray}
\dfrac{\partial}{\partial \rho} \left(\dfrac{\partial h(g,\rho)}{\partial g}\right) & = & \dfrac{- (\partial g / \partial \rho) (g^2 + \rho^2) + 4g(g ( \partial g / \partial \rho ) + \rho) }{(g^2 + \rho^2)^3} \left( 2\rho^2 + g^2(1 - Y(g,\rho)) \right) \nonumber \\
& & - \dfrac{g}{(g^2 + \rho^2)^2} \left(4\rho + 2g\dfrac{\partial g}{\partial \rho} \left( 1 - Y(g,\rho) \right) + g^2 \dfrac{g (\partial g / \partial \rho) + \rho}{Y(g,\rho)} \right) - \dfrac{g(\partial g / \partial \rho) + \rho}{(g^2 + \rho^2)^2} \left(2g(1 - Y(g,\rho)) +  \dfrac{g^3}{Y(g,\rho)} \right) \nonumber\\
& & + \dfrac{1}{2(g^2 + \rho^2)} \left( 2\dfrac{\partial g}{\partial \rho} (1 - Y(g,\rho)) + \dfrac{2g \rho}{Y(g,\rho)} + \dfrac{5g^2 (\partial g / \partial \rho) }{Y(g,\rho)} + g^3\dfrac{g(\partial g / \partial \rho) + \rho}{Y(g,\rho)^3} \right).
\end{eqnarray}
Finally,
\begin{eqnarray}
\dfrac{\partial q(\rho,u)}{\partial \rho} = \left(\dfrac{\partial P(\rho,u)}{\partial \rho}Q(\rho,u) - P(\rho,u)\dfrac{\partial Q(\rho,u)}{\partial \rho}\right) \Big/ Q(\rho,u)^2,
\end{eqnarray}
with
\begin{eqnarray}
\dfrac{\partial P(\rho,u)}{\partial \rho} & = & 
\left(3(1-\rho)(1+\rho)^2 - (1+\rho)^3 \right)u^2
\left[ 
\left( \dfrac{3\rho}{2} - 1 + \rho(1 + \rho)^3  u a_2(\rho, u) \right)
a_{12}(\rho) - \rho 
\left(1 + (1 + \rho)^3 u a_1(\rho,u)\right) 
a_{22}(\rho) \right] \nonumber \\
&&
+  (1-\rho)(1+\rho)^3 
u^2
\left[
\left(\dfrac{3}{2} + 3 u (1+\rho)^2 \rho a_2(\rho,u) + u(1+\rho)^3\left(a_2(\rho,u) + \rho \dfrac{\partial a_2(\rho,u)}{\partial \rho} \right) \right)a_{12}(\rho) \right. \nonumber \\
& & + \left.\left( \dfrac{3\rho}{2} - 1 + \rho(1+\rho)^3 u  a_2(\rho,u)\right) \dfrac{\partial a_{12}(\rho)}{\partial \rho} - \left(\rho \dfrac{\partial a_{22}(\rho)}{\partial \rho} + a_{22}(\rho)\right)\left( 1 + (1+\rho)^3 u a_1(\rho,u)\right) \right.\nonumber \\
&&\left. - \rho a_{22}(\rho)\left(3(1+\rho)^2 u a_1(\rho,u) + (1+\rho)^3 u \dfrac{\partial a_1(\rho,u)}{\partial \rho} \right) \right] \\
\dfrac{\partial Q(\rho,u)}{\partial \rho} & = & 2\dfrac{\partial g_0}{\partial \rho} \left(1+(1+\rho)^3 u a_2(\rho,u)\right)^2  + 4g_0\left(1+(1+\rho)^3 u a_2(\rho,u)\right) u \left(3(1+\rho)^2a_2(\rho,u) + (1+\rho)^3 \dfrac{\partial a_2(\rho,u)}{\partial \rho}\right). \nonumber \\
\end{eqnarray}
\end{widetext}

\newcommand{\Aa}[0]{Aa}
%

\end{document}